\documentclass[aps,prx,twocolumn,groupedaddress,superscriptaddress,longbibliography]{revtex4-2}

\usepackage{graphicx}
\usepackage{helvet}
\usepackage{amsmath,amssymb}
\usepackage{bm}
\usepackage{colortbl}
\usepackage{braket}
\begin{document}
\title{Volatile resistive-switched state in a bulk organic conductor with a sharp metal-insulator transition}

\author{Riku Ishii}
\affiliation{
Department of Applied Physics, Tokyo University of Science, Tokyo 125-8585, Japan}

\author{Ryo Motohashi}
\affiliation{
Department of Applied Physics, Tokyo University of Science, Tokyo 125-8585, Japan}

\author{Keitaro Tada}
\affiliation{
Department of Applied Physics, Tokyo University of Science, Tokyo 125-8585, Japan}

\author{Yusuke Suzuki}
\affiliation{
Department of Applied Physics, Tokyo University of Science, Tokyo 125-8585, Japan}

\author{Takayoshi Kouchi}
\affiliation{
Department of Applied Physics, Tokyo University of Science, Tokyo 125-8585, Japan}

\author{Hiroshi Oike}
\affiliation{
Research Center for Materials Nanoarchitechtonics (MANA), National Institute for Materials Science (NIMS), Tsukuba-shi, Ibaraki 305-0047, Japan}
\affiliation{
PRESTO, Japan Science and Technology Agency (JST), Kawaguchi-shi, Saitama 332-0012, Japan}

\author{Fumitaka Kagawa}
\affiliation{
RIKEN Center for Emergent Matter Science (CEMS), Wako-shi, Saitama 351-0198, Japan}
\affiliation{
Department of Physics, Institute of Science Tokyo, Tokyo 152-8551, Japan}

\author{Reizo Kato}
\affiliation{
RIKEN, Cluster for Pioneering Research (RIKEN-CPR), Wako-shi, Saitama 351-0198, Japan}

\author{Tetsuaki Itou}
\email{tetsuaki.itou@rs.tus.ac.jp}
\affiliation{
Department of Applied Physics, Tokyo University of Science, Tokyo 125-8585, Japan}

\date{August 20, 2026}

\begin{abstract}

	Volatile resistive switching in correlated-electron systems, characterized by an abrupt resistance decrease under applied current, is crucial for developing next-generation electronics.
	Despite its technological significance, the underlying physics remains elusive.
	Inorganic thin films on substrates---the widely studied platform for resistive switching---usually exhibit broad temperature-induced metal-insulator transitions (MITs) and substantial heat dissipation.
	These factors complicate the nonlinear thermal effect induced by Joule heating, a key contributor to resistive switching, rendering it excessively complex and difficult to decipher.
	Here we investigate a resistive-switched state in the \textit{bulk} organic conductor ($d$7-DMe-DCNQI)$_{2}$Cu, which undergoes an extremely sharp first-order MIT and exhibits weak heat dissipation, using resistance and $^{1}$H-NMR measurements.
	These extreme conditions make the Joule heating effect vivid, allowing us to observe peculiar phenomena, including temperature locking to the MIT and `inverse Ohm's law'---an \textit{inverse} proportionality between voltage and current.
	These findings provide fundamental insights into the nonlinear thermal effect in resistive switching, offering a pathway to efficient resistive-switching technologies.

\end{abstract}

\maketitle

\section{Introduction}
	Metal-insulator transitions (MITs) in electron systems, induced by electronic correlations and often intertwined by low dimensionality, have been central issues in fundamental physics.
	Over the course of a century of extensive research, both theoretical and experimental physicists have established a solid foundational understanding of these MITs---namely, the Mott and Peierls transitions---typically triggered by changes in temperature or pressure~\cite{Imada1998}.
	In the insulating phases near the transitions, it has also been empirically observed that applying an electric field/current can sometimes cause a significant drop in resistance, almost certainly related to the transitions, a phenomenon known as `resistive switching'.
	This phenomenon is the focus of intense research in applied physics, because it holds promise as a candidate for various innovative applications, such as resistive memories~\cite{Son2011,Chang2011,Pellegrino2012,Bae2013,Janod2015}, optoelectronics ~\cite{Coy2010,Liu2012,Markov2015,Fan2016,Lee2017,Miller2017,Butakov2018,Sanchez2018,Olivares2018,Butakov2018_2,Coll2019,Cueff2020,Li2022} and neuromorphic computing~\cite{Pickett2013,Ignatov2015,Moon2015,YouZhou2015,Prezioso2015,Kumar2017,Stoliar2017,Yi2018,delValle2018,Lin2018,Bohaichuk2019,Salev2019,Feldmann2019,delValle2020,Kumar2020,Oh2021,Lin2023,Schofield2023}.
	
	Despite the intensive interest and numerous experimental studies, however, the fundamental understanding of the mechanism behind volatile resistive switching remains incomplete over the long term.
	Although Joule heating has been considered to play invariably a significant role, it is sometimes nontrivial whether resistive switching is solely triggered by Joule heating~\cite{Zimmers2013,Shukla2014,Liu2016,delValle2021,Salev2021,Rocco2022,Adda2022,Luibrand2023,Rischau2024} or influenced also by electronic effects~\cite{Valmianski2018,Kalcheim2020,Sahoo2023}.
	One reason for this uncertainty is that resistive switching has been studied mainly in inorganic thin films with complex parameters, which complicate the Joule heating effect, making it difficult to accurately interpret experimental observations in such systems.
	For example, while the sharpness of the MIT has been identified as a critical parameter in resistive switching caused by Joule heating~\cite{delValle2021,Rocco2022,Luibrand2023,Rischau2024,delValle2021_2}, most of the inorganic thin films studied so far exhibit resistance ratios ($R_{\mathrm{ins}} / R_{\mathrm{metal}}$) of no more than five orders of magnitude, as well as distributions of the transition temperature ($\Delta T_{\mathrm{c}}$)~\cite{Zimmers2013,Shukla2014,delValle2021,Salev2021,Rocco2022,Adda2022,Luibrand2023,Rischau2024,Kalcheim2020,delValle2021_2,Brockman2014,Huang2014,delValle2019}.
	In addition, it has been proposed that smaller heat dissipation makes resistive switching easier to achieve~\cite{Rocco2022,Rischau2024}, and also that excessively strong heat dissipation makes the switching phenomena stochastic and thus unpredictable~\cite{Rocco2022}.
	Therefore, it is desired to study the switching phenomena in systems with low heat dissipation.
	However, most investigations to date have been conducted on thin films on substrates, where strong dissipation to the substrate is inevitable.

\begin{figure*}
\includegraphics[width=0.8\textwidth]{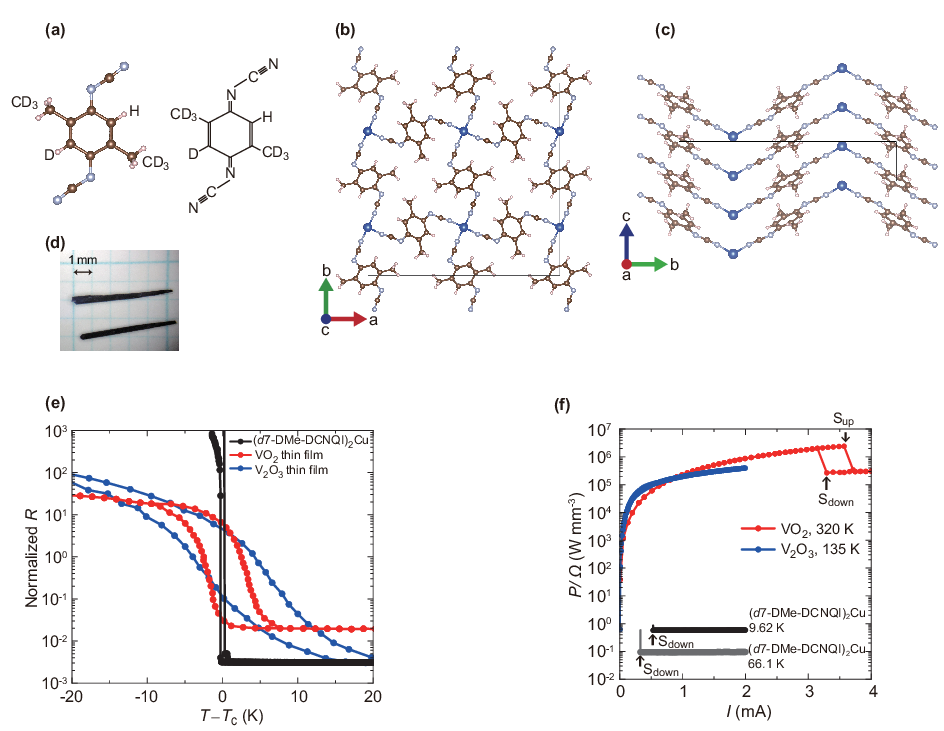}
\caption{Structure and transport properties of ($d$7-DMe-DCNQI)$_{2}$Cu. (a) Molecular structure of ($d$7-DMe-DCNQI)$_{2}$Cu. The $d$7 means that the seven $^{1}$H sites are substituted with deuterons. (b) ($d$7-DMe-DCNQI)$_{2}$Cu structure viewed down the $c$-axis. The space group is $I4_{1}$/a, and the structure consists of planar DCNQI molecules stacked along the $c$-axis. (c) ($d$7-DMe-DCNQI)$_{2}$Cu structure viewed down the $a$-axis. In the insulating phase, the threefold superlattice occurs along the stacking axis. The charge ordering at the Cu sites follows a Cu$^{+}$ Cu$^{+}$ Cu$^{2+}$ pattern. (d) Photograph of needlelike crystals of ($d$7-DMe-DCNQI)$_{2}$Cu. (e) $T-T_{\mathrm{c}}$ dependence of the normalized resistance for VO$_{2}$ and V$_{2}$O$_{3}$ thin films and ($d$7-DMe-DCNQI)$_{2}$Cu. The normalized resistance represents the sample resistance divided by the center value of the hysteresis loop, and $T-T_{\mathrm{c}}$ denotes the difference between the sample temperature and the hysteresis midpoint temperature. (f) Current dependence of power ($I \times V$) divided by volume, representing Joule heating per unit volume, for VO$_{2}$ and V$_{2}$O$_{3}$ thin films and ($d$7-DMe-DCNQI)$_{2}$Cu. S$_{\mathrm{down}}$ and S$_{\mathrm{up}}$ represent the switching points when the current is decreased and increased, respectively. The data in panels (e) and (f) for the VO$_{2}$ and V$_{2}$O$_{3}$ thin films were obtained by extracting values from Ref.~\cite {delValle2021_2}.
}
\end{figure*}

	Under these circumstances, we present an experimental elucidation of the volatile resistive-switched state induced by an applied electric current in a \textit{bulk} system exhibiting an \textit{extremely sharp} MIT, aiming to establish a robust foundation for understanding resistive switching caused by Joule heating and to re-evaluate the underlying behavior of the resistive-switched state under a sharp MIT and low heat dissipation.
	The material of focus is the quasi-one-dimensional organic conductor ($d$7-DMe-DCNQI)$_{2}$Cu (Fig.~1(a)-(d)), which exhibits an extremely sharp 1st-order MIT at 79 K~\cite{Aonuma1993,Aonuma1995,Kato2000}.
	In addition, the sample used in this study is a bulk crystal rather than a thin film on a substrate.
	
	The family of DCNQI (N,N$^{\prime}$-dicyanoquinonediimine) compounds provides a wide platform for studying such MITs in organic conductors. Their electronic states are highly tunable by chemical substitution or external pressure. For instance, the non-deuterated (DMe-DCNQI)$_{2}$Cu remains metallic down to low temperatures, whereas deuteration of the DCNQI molecule or the application of slight pressure induces a sharp first-order MIT~\cite{Aonuma1993,Aonuma1995,Kato2000}, which is considered to be caused by a synergetic mechanism involving both the Mott and Peierls transition natures~\cite{Fukuyama1992,Fukuyama2006}. This sensitivity to isotopic substitution and pressure reflects the delicate balance between electron correlation and lattice effects in this molecular system, making the DCNQI family an ideal framework for exploring correlation-driven transitions. In this study, we focused on ($d$7-DMe-DCNQI)$_{2}$Cu, in which seven of the eight $^{1}$H sites in the DCNQI molecule are replaced by deuterons, as shown in Fig.~1(a).
	
	The sample in this study exhibits a metallic phase above 79 K with a two-wire resistance of $\sim$10 $\Omega$, and an insulating phase below 78 K, where the resistance exceeds the measurement limit of 1 M$\Omega$, showing an enormous ratio $R_{\mathrm{ins}} / R_{\mathrm{metal}} > 10^{5}$.
	More importantly, the transitions on heating and cooling occur within $\pm 0.5$ K temperature range, and the hysteresis loop is approximately $\pm1$ K (Fig.~1(e)).
	This sharp transition contrasts with transitions observed in inorganic thin films in which resistive switching has been studied so far.
	For example, VO$_{2}$ and V$_{2}$O$_{3}$ thin films typically have a transition temperature distribution $\Delta T_{\mathrm{c}}$ ranging between $\pm20$ K~\cite{Zimmers2013,Rocco2022,Rischau2024,Kalcheim2020,delValle2021_2,Brockman2014,delValle2019}.
	A more significant point is that the sample in this study is a \textit{bulk} crystal without a substrate.
	In this situation, heat dissipation outside the sample occurs through the surrounding He gas for temperature control and through the wiring for electric current application.
	Consequently, the dissipation is significantly lower, much weaker than that in thin films on substrates.
	Indeed, Joule heating per unit volume in the resistive-switched state of this study, which balances heat dissipation, is more than five orders of magnitude smaller than that in the resistive-switched states of VO$_{2}$ and V$_{2}$O$_{3}$ thin films, as shown in Fig.~1(f).
	The present study seeks a paradigm shift to uncover phenomena occurring under low heat generation/dissipation, in which resistive switching takes place under minimal current density and the Joule heating is unequivocally responsible for the resistive switching.
	
	From the perspective of fundamental physics, the low-resistance state under an electric current after resistive switching is nothing but a `dissipative structure', where the strong nonlinearity in the relationships between flows and forces (corresponding to negative differential resistance in the context of resistive switching) induces spinodal-like decomposition in nonequilibrium steady states~\cite{Ridley1963,Kumar2018,Goodwill2019}.
	In the field of resistive switching, this phenomenon has been discussed in terms of spontaneous `filament configuration' or `current constriction', where metallic current filament forms within an insulating background~\cite{Zimmers2013,Liu2016,delValle2021,Salev2021,Rocco2022,Adda2022,Luibrand2023,Rischau2024,Kalcheim2020,delValle2021_2,Brockman2014,delValle2019,Kumar2018,Goodwill2019,Kumar2017_2}.
	We demonstrate that such spatial self-organization is indeed realized even in the present \textit{bulk} sample under applied electric current, using NMR techniques, which can provide microscopic information throughout the \textit{bulk} sample and also can measure the \textit{internal} temperature of the sample.

\section{Results}

\subsection{Experimental setup}

	Due to the sharp MIT and low heat dissipation in the present sample, the spatial self-organized state can be achieved with a relatively small current density.
	Consequently, the self-organized state is stabilized in the present sample down to an ambient temperature of absolute zero with an application of only a 2.0 mA current (corresponding to a current density of $\sim$10 A/cm$^{2}$).
	This small current density ensures that the electronic effect is not active and that the resistive switching is governed by the Joule heating mechanism.
	Note that, due to the nearly infinite $R_{\mathrm{ins}}$, this state cannot be realized experimentally by applying an electric voltage (up to 210 V in our experimental setup) to the insulating state.
	Instead, we achieved the state by cooling the sample with a constant current applied, from temperatures above the transition temperature of 79 K, where the metallic state is realized.
	We later show through microscopic NMR results that the state obtained in this process is indeed a spatially self-organized state, which can be considered a resistive-switched state, similar to those obtained by applying an electric voltage under constant ambient temperatures in previous studies.

\begin{figure}
\includegraphics[width=0.8\hsize]{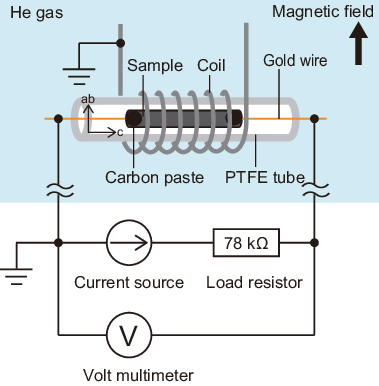}
\caption{Schematic of the experimental setup for resistance and $^{1}$H-NMR measurements on a bulk single crystal of ($d$7-DMe-DCNQI)$_{2}$Cu. The crystal dimensions are $0.13 \times 0.13 \times 6.3$ mm$^{3}$. The sample was placed in a Teflon tube, around which an NMR coil was wound, and a magnetic field of 2.62 T was applied approximately perpendicular to the $c$-axis. To apply an electric current, two gold wires ($\phi$ 15 $\mu$m) were attached to the crystal edges using carbon paste. The measurements were performed using a current source (Keithley 2450) with a 78 k$\Omega$ resistor to stabilize the negative differential resistance state.}
\end{figure}

	Figure~2 illustrates the setup for resistance and $^{1}$H-NMR measurements on a bulk needlelike single crystal of ($d$7-DMe-DCNQI)$_{2}$Cu (see the Appendix for details).
	In this configuration, the sample was not mounted on a substrate, and heat dissipation occurred through the gold wires and He gas ($\sim$ 0.7 atm) filling the sample chamber, resulting in minimal heat dissipation.
	The ambient temperature was measured using a resistance sensor (Cernox), located a few centimeters from the sample.

\subsection{Resistance under applied current}

	Figure~3(a) shows the ambient temperature dependence of the two-wire sample resistance (defined as sample voltage divided by applied current, $V/I$) under the infinitesimal current limit (equilibrium conditions, 0.001 mA and 0.01 mA below and above the transition temperature, respectively) and with constant currents applied (current values: 0.3 mA, 0.5 mA, and 2.0 mA).
	The data under equilibrium conditions were collected during both heating and cooling processes, while the data under constant currents were obtained only during cooling process.

\begin{figure}
\includegraphics[width=\hsize]{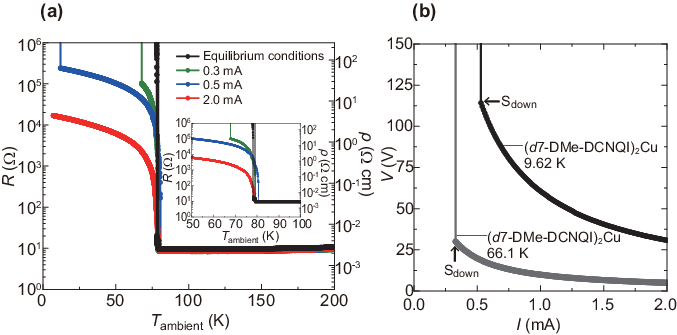}
\caption{Transport properties of ($d$7-DMe-DCNQI)$_{2}$Cu. (a) Ambient temperature dependence of the resistance of ($d$7-DMe-DCNQI)$_{2}$Cu. The black plot shows the two-wire resistance under equilibrium conditions, while the other plots show the two-wire resistance under applied constant currents (0.3 mA, 0.5 mA, 2.0 mA). The inset shows the behavior of the resistance around the transition temperature. The data under equilibrium conditions are the same as the black data shown in Fig.~1(e). (b) Two-probe $I$(current)--$V$(voltage) characteristics at 9.62 K and 66.1 K for ($d$7-DMe-DCNQI)$_{2}$Cu. S$_{\mathrm{down}}$ indicates the switching points when the current is decreased. These data yield the current dependence of power ($I \times V$) per unit volume shown in Fig.~1(f).}
\end{figure}

	Under equilibrium conditions, a sharp and significant resistive jump was observed, as reported in earlier studies~\cite{Aonuma1993,Aonuma1995,Kato2000}.
	As stated in the introduction, the resistance ratio between a metal and an insulator under equilibrium conditions is generally regarded as a critical parameter in resistive switching.
	As shown in Fig.~1(e), in the present system ($d$7-DMe-DCNQI)$_{2}$Cu, the resistance ratio is notably large, and the transition is much sharper compared to those in VO$_{2}$ and V$_{2}$O$_{3}$ thin films. (We note that the comparison in Fig.~1 is made with thin-film data of VO$_{2}$ and V$_{2}$O$_{3}$, whose transitions are generally broader than those of bulk single crystals. Single-crystal VO$_{2}$ and V$_{2}$O$_{3}$ exhibit much sharper transitions~\cite{NFMott1990}.)

	The data under constant currents exhibit gradual increases in resistance below the transition temperature (79 K), which sharply contrasts to the high resistance in the insulating state under equilibrium conditions.
	The resistance then abruptly switches to the insulating state at 67.8 K under 0.3 mA, and 12.0 K under 0.5 mA.
	Notably, the intermediate resistance is stabilized down to zero ambient temperature under 2.0 mA.
	This intermediate resistance indicates that a non-equilibrium steady state is realized in the sample under these currents, analogous to the resistive-switched states discussed in prior studies investigating the effects of current/voltage application under constant ambient temperatures.
	Indeed, when the current is reduced under a constant ambient temperature, drastic resistive switching behavior is observed in the $I$(current)--$V$(voltage) characteristics, as shown in Fig.~3(b).
	Note that, once the resistance switches to the insulating state, the intermediate resistance state cannot be re-established by increasing the voltage, even up to 210 V.
	However, after heating the sample above the transition temperature and subsequently cooling it down following the same procedure, the same intermediate resistance state can be reproduced.
	In the present sample geometry, the maximum applied voltage of 210 V corresponds to an electric field of $\sim$300 V/cm, under which switching from the insulating state back to the intermediate resistance state could not be achieved.
	Reducing the crystal size allows access to higher electric fields and may enable switching from the insulating to the intermediate-resistance state.

\subsection{$^{1}$H-NMR under applied current}

	To discuss whether the intermediate resistance state under current is indeed a spatially self-organized state that can be considered a volatile resistive-switched state, we performed $^{1}$H-NMR measurements, which provide microscopic information on the bulk sample.

\begin{figure}
\includegraphics[width=\hsize]{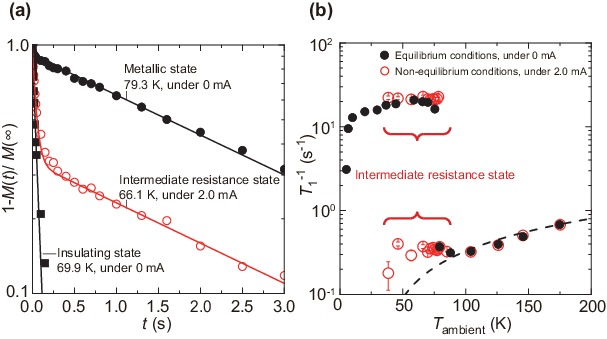}
\caption{$^{1}$H-NMR relaxation curves and relaxation rate ($T_{1}^{-1}$) of ($d$7-DMe-DCNQI)$_{2}$Cu. (a) $^{1}$H-NMR relaxation curves for the metallic and insulating states under 0 mA, and the intermediate resistance state under 2.0 mA. The black solid curves represent fits to single-exponential functions for the metallic and insulating states, while the red solid curve represents a fit to a double-exponential function for the intermediate resistance state under 2.0 mA. The excellent fit of the double-exponential function provides evidence for the coexistence of the metallic and insulating phases in the intermediate resistance state. (b) Ambient temperature dependence of $T_{1}^{-1}$ under 0 mA (equilibrium conditions) and under 2.0 mA (intermediate resistance state). $T_{1}^{-1}$ under 2.0 mA is decomposed into two relaxation components via the double-exponential fitting shown in (a). The dashed black line indicates the Korringa relation ($T_{1}^{-1} \propto T_{\mathrm{ambient}}$) expected for the metallic state under equilibrium conditions.}
\end{figure}

	Figure~4(a) illustrates the $^{1}$H-NMR relaxation curve for the intermediate resistance state realized under 2.0 mA at 66.1 K.
	For comparative analysis, the figure also presents the relaxation curves for the metallic (79.3 K) and insulating (69.9 K) phases under equilibrium conditions.
	In the metallic and insulating phases, the relaxation curves exhibit nearly single-exponential behaviors, ensuring that the system is indeed in a single uniform phase.
	The relaxation time ($T_{1}$) in the insulating phase is more than an order of magnitude faster than that in the metallic phase, which is due to localized spins associated with the Mott nature~\cite{Hiraki1995}.
	In the intermediate resistance state, contrastingly, the relaxation curve is not a single exponential but is described by a double-exponential function, wherein the slow and fast relaxation times align with those of the metallic and insulating phases, respectively.
	This result suggests that, in the intermediate resistance state, spatial coexistence of the metallic and insulating phases occurs.
	Figure~4(b) illustrates the ambient temperature dependence of the fast and slow relaxation rates ($1/T_{1}$) in the intermediate resistance state realized under 2.0 mA, in comparison to those in the metallic and insulating phases under equilibrium conditions.
	The two values of $1/T_{1}$ in the intermediate resistance state are virtually independent of ambient temperature over the entire temperature range, corresponding to the $1/T_{1}$ values for the metallic and insulating phases near the transition temperature.
	
\begin{figure}
\includegraphics[width=\hsize]{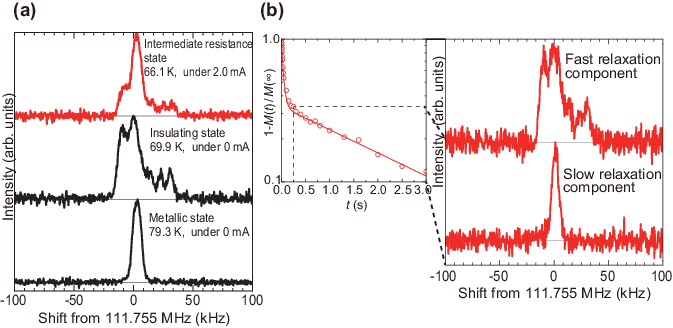}
\caption{$^{1}$H-NMR spectra of ($d$7-DMe-DCNQI)$_{2}$Cu. (a) $^{1}$H-NMR spectra of the metallic and insulating states under 0 mA, and the intermediate resistance state under 2.0 mA. (b) Decomposed $^{1}$H-NMR spectra of the intermediate resistance state. The fast relaxation component corresponds to the spectrum that recovers within 0.25 s after the saturation comb pulses, while the slow relaxation component corresponds to the spectrum that recovers after 0.25 s.}
\end{figure}

	Figure~5(a) shows the $^{1}$H-NMR spectra of the intermediate resistance state under 2.0 mA, as well as those of the metallic and insulating phases (detailed data on the ambient temperature dependence of the spectra are presented in the Supplemental Material~\cite{Supplementary}).
	The spectrum of the intermediate resistance state combines characteristics of both the metallic and insulating phases.
	As shown in Fig.~5(b), the fast relaxation component matches that of the insulating phase, whereas the slow relaxation component coincides with the metallic phase.
	This result further provides clear support for the spatial coexistence of the metallic and insulating phases in the intermediate resistance state under current; taking into account the intermediate resistance, metallic current filament forms within an insulating background.

\subsection{Effect of the Joule heating}

	To investigate the effect of Joule heating, we measured the sample temperature influenced by the applied current as a function of ambient temperature.
	The sample temperature was estimated not through optical surface imaging, often used in thin-film research, but rather via $^{1}$H-NMR signal intensity, which reflects the average temperature across the bulk sample, as it is essential to acquire temperature information for the entire bulk.

\begin{figure}
\includegraphics[width=0.8\hsize]{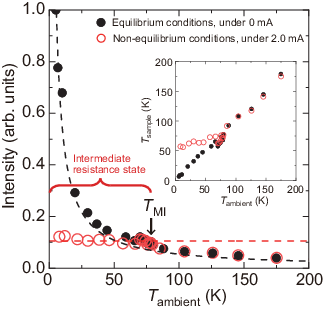}
\caption{Ambient temperature dependence of the $^{1}$H-NMR signal intensity, which is inversely proportional to the sample temperature. The black closed circles represent the intensity under 0 mA, while the open red circles represent the intensity under 2.0 mA. The intensity under 0 mA exhibits inverse proportionality with ambient temperature (black dashed line), demonstrating that the sample temperature is equal to the ambient temperature. In the region where $T_{\mathrm{ambient}} > 79$ K ($T_{\mathrm{MI}}$), the intensity under 2.0 mA is the same as that under 0 mA, indicating that the sample temperature is equal to the ambient temperature even under 2.0 mA. However, it remains nearly constant in the region where $T_{\mathrm{ambient}} < 79$ K (red dashed line), demonstrating that the sample temperature in the intermediate resistance state is kept nearly constant. The inset shows the ambient temperature dependence of the sample temperature, estimated from the NMR signal intensity.}
\end{figure}

	Figure~6 illustrates the ambient temperature dependence of the $^{1}$H-NMR signal intensity, defined as the integral of the Fourier-transform spectrum over the entire frequency range.
	In general, NMR signal intensity is proportional to the inverse of the sample temperature, as described by Curie's law, independent of whether the system is metallic or insulating. Thus, the NMR intensity can serve as a probe of the sample temperature.
	As shown in the figure, the intensity under equilibrium conditions is proportional to the inverse of the ambient temperature.
	This result demonstrates that the sample temperature under equilibrium conditions is identical to the ambient temperature, as expected.
	Above the transition temperature (79 K), the intensity under a current of 2.0 mA also follows this proportionality, indicating negligible Joule heating.
	However, below the transition temperature, the intensity remains almost constant, regardless of the cooling of the ambient temperature.
	This result clearly demonstrates that, in the intermediate resistance state, the average sample temperature is significantly increased from the ambient temperature by Joule heating.
	The average sample temperature, though slightly reduced at low ambient temperatures (details are discussed in the Supplemental Material~\cite{Supplementary}), is almost locked to the transition temperature, as shown in the inset of Fig.~6.

\subsection{Applied current dependence of the sample temperature}

	To explore phenomena caused by the temperature locking effect discussed in the previous section, we return to the transport properties.
	Figure~7(a) shows the $I$--$V$ characteristics of the intermediate resistance state at several ambient temperatures, measured by decreasing the applied current.
	At each ambient temperature, the intermediate resistance state exhibits negative differential resistance and switches to the insulating state at low current.
	
\begin{figure}
\includegraphics[width=\hsize]{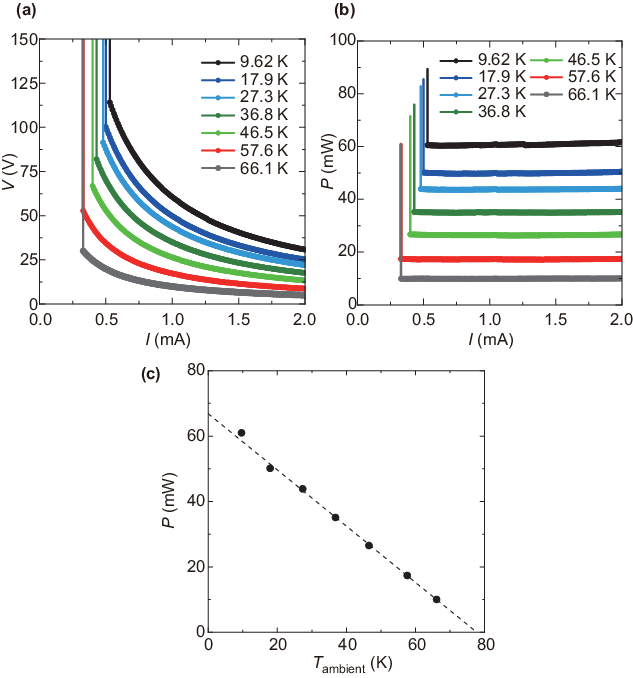}
\caption{Transport characteristics at different ambient temperatures. (a) $I$(current)--$V$(voltage) characteristics at different ambient temperatures. The data were measured by decreasing the current in 0.01 mA steps. The abrupt increases in voltage upon decreasing the current correspond to switching to the insulating state. (b) $I$--$P$ (calculated as $I \times V$) characteristics at different ambient temperatures. (c) Ambient temperature dependence of $P$ in the intermediate resistance state. The dashed line serves as a guide to the eye, representing $P \propto ($79 K$ - T_{\mathrm{ambient}})$}
\end{figure}

	In the intermediate resistance state, remarkably, the voltage and current exhibit an inverse proportional relationship, $V\propto 1/I$, which can be termed `inverse Ohm's law'.
	This is explicitly shown in Fig.~7(b), where the power ($I \times V$) remains constant in this region, regardless of the applied current.
	Here, we discuss the underlying mechanism behind `inverse Ohm's law'.
	When the sample temperature is locked, heat dissipation outside the sample---determined by the difference between the sample and ambient temperatures---remains invariant as long as the ambient temperature is unchanged, regardless of the applied current.
	In addition, since the intermediate resistance state is a steady state, the following relationship holds: (Joule heating) = (Heat dissipation).
	Thus, in the state with the sample temperature locked, Joule heating, which balances with heat dissipation, must remain at a fixed value to maintain the sample temperature, independent of the applied current.
	This scenario explains the observation of `inverse Ohm's law' shown in Fig.~7. 
	In this regime, the current filament thickens with increasing applied current and thins with decreasing current, as required to maintain a nearly constant level of Joule heating.

	Figure~7(c) shows the ambient temperature dependence of $P (= I \times V)$ in the intermediate resistance state.
	When the ambient temperature is decreased, the heat dissipation increases owing to the increase in the difference between the ambient temperature and the sample temperature (almost fixed around the transition temperature of 79 K), and consequently, Joule heating required to achieve the intermediate resistance state inevitably increases.
	This explains why the heat generation ($P = I \times V$), or the proportionality factor of the `inverse Ohm's law', systematically increases as the ambient temperature decreases, following $P \propto ($79 K$ - T_{\mathrm{ambient}})$, as shown in Fig.~7(c).
	This increase in the proportionality factor with cooling is most likely caused by the thinning of the current filament.
	Figure~8 illustrates the ambient temperature dependence of the volume fraction of the metallic phase in the intermediate resistance state under an applied current of 2.0 mA, estimated from the fraction of the slow relaxation component in the $^{1}$H-NMR relaxation curves.
	This figure demonstrates that decreasing the ambient temperature reduces the volume fraction of the metallic phase, consistent with the scenario of current filament thinning.

\begin{figure}
\includegraphics[width=0.6\hsize]{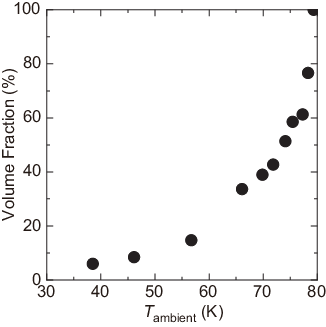}
\caption{Ambient temperature dependence of the volume fraction of the metallic phase in the intermediate resistance state under an applied current of 2.0 mA. The volume fraction is estimated from the fraction of the slow relaxation component in the $^{1}$H-NMR relaxation curves.}
\end{figure}

\section{Discussion}

	As discussed in Fig.~6, the $^{1}$H-NMR intensity remains almost constant in the intermediate resistance state. We interpret this nearly constant NMR intensity as evidence that the average sample temperature is almost locked around the transition temperature. Note that the NMR intensity reflects the spatially averaged temperature of the entire sample, and thus local variations may exist---for example, the metallic filament could be hotter than the surrounding insulating regions. However, as discussed below, such temperature gradients are expected to remain modest in the present system because of the extremely weak heat generation and dissipation per unit volume. (In this sense, the ``temperature locking'' refers to the locking of the average sample temperature, rather than complete thermal uniformity throughout the sample.)
	
	We propose that this ``temperature locking'' arises from the two extreme conditions of the present system, as detailed below.
	One condition is the intense and sharp MIT---characterized by a high $R_{\mathrm{ins}} / R_{\mathrm{metal}} > 10^{5}$ within a narrow transition temperature range of $\sim$1 K.
	This ensures that, in the intermediate resistance state, where the metallic and insulating phases coexist, the temperature at the well-defined boundaries between the two phases should be precisely maintained at the transition temperature of 79 K.
	The other condition is weak heat dissipation outside the sample, thereby resulting in weak heat generation (Joule heating) per unit volume.
	In principle, the temperature in the metallic phase, where Joule heating occurs due to current concentration, should be higher than the boundary temperature (79 K), while the temperature in the insulating phase, where no Joule heating occurs, should be lower than the boundary temperature.
	However, in the present bulk system in a He gas environment, the thermal coupling to the surroundings is expected to be much weaker than that within the sample.
	Thus, the heat transfer process ``from the insulating background to the surroundings'' is much less efficient than the process ``from the metallic filament to the insulating background''. 
	This results in minimal temperature variations across the sample.
	Therefore, the temperatures in the metallic and insulating regions do not deviate significantly from the boundary temperature of 79 K, as demonstrated by the nearly constant NMR intensity shown in Fig.~6. (Indeed, the metallic and insulating $1/T_{1}$ values in the intermediate resistance state shown in Fig.~4 appear almost constant as a function of the ambient temperature. Because the intrinsic temperature dependence of the metallic $1/T_{1}$ should obey the Korringa relation ($1/T_{1} \propto T_{\mathrm{sample}}$), the nearly constant value of the metallic $1/T_{1}$ implies that the metallic filament is not heated far above the transition temperature.)
	These two characteristics keep the average sample temperature in the intermediate resistance state close to the transition temperature of 79 K, as shown in the right panel of Fig.~9(b). Note that under these small temperature variations, the sharp MIT in the present system ensures drastic current-density concentration, as shown in the left panel of Fig.~9(b).
	This temperature locking stands in striking contrast to the wide temperature distribution (Fig.~9(a)) expected in conventional inorganic thin films with broad MITs and substantial heat dissipation to substrates.
	As discussed previously, this distinct thermal behavior gives rise to the fundamental physical phenomenon of `inverse Ohm's law'.
	Our finding of temperature locking also provides a strategy for developing durable resistive switching devices---namely, a directional approach to prevent local overheating, which can sometimes destroy them.

\begin{figure}
\includegraphics[width=\hsize]{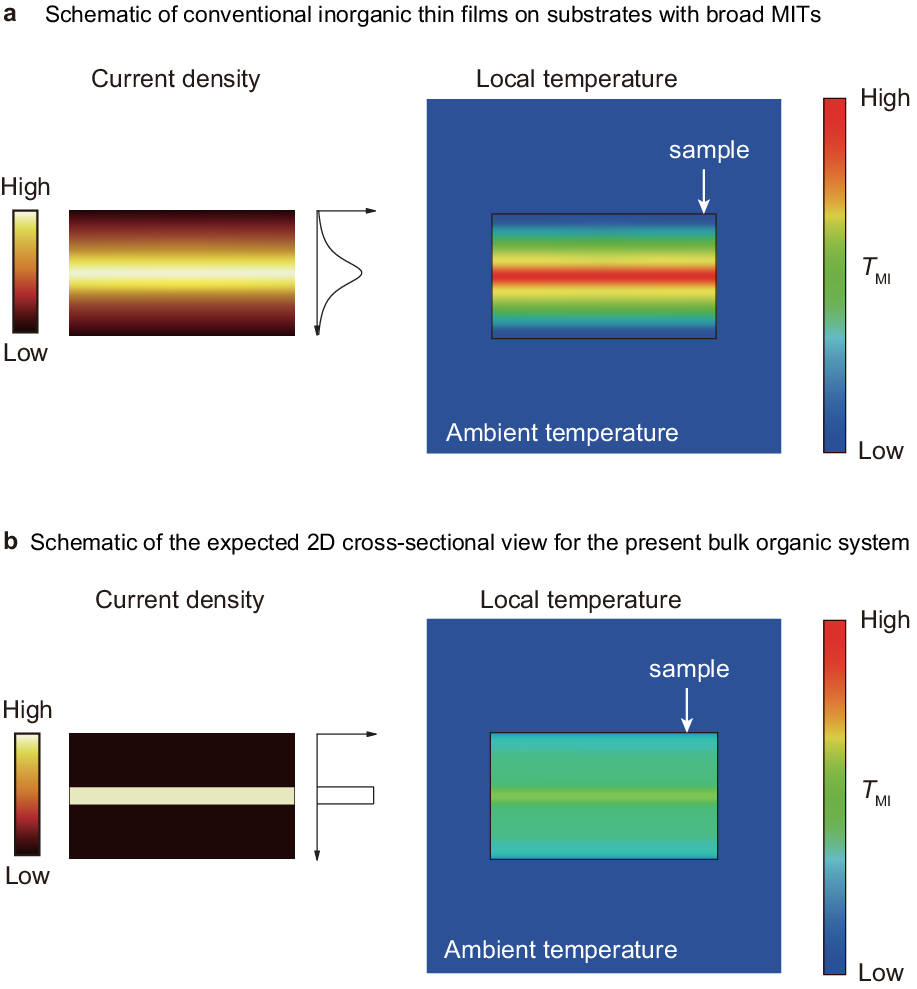}
\caption{Schematic figures of the current density and local temperature distributions for a current filament in volatile resistive-switched states dominated by Joule heating. (a) Schematic for conventional inorganic thin films on substrates with broad MITs and substantial heat dissipation to the substrates. (b) Schematic of the expected 2D cross-sectional view for the present bulk organic system with a sharp MIT and low heat dissipation outside the sample.}
\end{figure}

	In summary, we clarified the nature of the intermediate resistance state induced by applied current---regarded as a volatile resistive-switched state---in a bulk organic MIT system under extreme conditions.
	This clarification was achieved through resistance measurements and $^{1}$H-NMR measurements that provide information on the bulk state across the sample.
	The extreme conditions---the intenseness/sharpness of the MIT and weak heat generation/dissipation---lead to two effects: (1) locking the sample temperature to the transition temperature and (2) `inverse Ohm's law,' an inverse proportional relationship between $V$ and $I$.
	These findings provide a solid foundation for understanding resistive switching driven by Joule heating and pave the way for developing more efficient resistive switching technologies.

\begin{acknowledgments}
	The authors thank T. Fusamae, Y. Sonobe, T. Sekizawa, and H. Mori for experimental assistance and fruitful discussions at the early stage of this project.
	This work was supported in part by KAKENHI (Grant Nos. 22H01184, 19H01852, and 18H05225) from Japan Society for the Promotion of Science (JSPS), and Japan Science and Technology Agency (JST) CREST (Grant No. JPMJCR23A1), Japan.
\end{acknowledgments}

\vspace{\baselineskip}

\appendix*
\section{Methods}
\textbf{Sample preparation.}
	The single crystals of ($d$7-DMe-DCNQI)$_{2}$Cu were prepared from $d$7-deuterated DMe-DCNQI and CuI by the diffusion method in an acetonitrile (CH$_{3}$CN) solution.
	A number of fine needlelike crystals, with typical dimensions of $0.1 \times 0.1 \times 5$ mm$^{3}$, were obtained, as shown in Fig.~1(d).

\vspace{\baselineskip}

\textbf{Details of experimental setup for resistance and $^{1}$H-NMR measurements.}
	Figure~2 illustrates the setup for resistance and $^{1}$H-NMR measurements on a bulk needlelike single crystal of ($d$7-DMe-DCNQI)$_{2}$Cu.
	The crystal has dimensions of $0.13 \times 0.13 \times 6.3$ mm$^{3}$, and both measurements were performed on the same crystal using the same experimental setup.
	To perform the $^{1}$H-NMR measurements, the sample was placed in a Teflon tube, around which an NMR coil was wound.
	A magnetic field of 2.62 T was applied approximately perpendicular to the $c$-axis.
	To apply an electric current, two gold wires ($\phi$ 15 $\mu$m) were attached to the crystal edges using carbon paste.
	The current was supplied using a constant current source (Keithley 2450) with a voltage limit of 210 V, and a 78 k$\Omega$ resistor was connected in series with the sample to stabilize the negative differential resistance state.

\vspace{\baselineskip}

\textbf{$^{1}$H-NMR measurements.}
	The $^{1}$H-NMR measurements were conducted at a magnetic field of 2.62 T, within an ambient temperature range of 7.72--205 K.
	The $^{1}$H-NMR spectra were obtained by the Fourier transformation of the spin-echo signals following the $\pi/2-\pi$ rf pulse sequence.
	The widths of the $\pi/2$ and $\pi$ pulses were typically 0.6 $\mu$s and 1.2 $\mu$s, respectively.
	The signal intensity was estimated by integrating the Fourier-transformed signal over the entire frequency range.
	The spin-lattice relaxation rates were determined from the recovery curves of the signal intensity following the irradiation of the comb pulses.

%

%

\clearpage
\onecolumngrid

\setcounter{figure}{0}
\renewcommand{\thefigure}{S\arabic{figure}}

\begin{center}
{\large\bfseries Supplemental Material for}\\[0.8ex]
{\large\bfseries
``Volatile resistive-switched state in a bulk organic conductor with a sharp metal-insulator transition''}

\vspace{1.5\baselineskip}

\end{center}

\vspace{2\baselineskip}

\section*{$^{1}$H-NMR spectra}

Figure~S1 shows the ambient temperature dependence of the $^{1}$H-NMR spectra of ($d$7-DMe-DCNQI)$_{2}$Cu at an applied magnetic field of 2.62 T, under equilibrium conditions without an applied current and non-equilibrium conditions with an applied current of 2.0 mA.
The experimental setup is described in the Appendix of the main text.

Under equilibrium conditions (Fig.~S1(a)), single peak spectra were observed in the metallic phase above 79.3 K because the magnetization induced by the applied magnetic field, and consequently the internal magnetic fields on the $^{1}$H nuclei caused by the magnetization, are negligibly small in the metallic phase.
By contrast, in the insulating phase below 78.3 K, the spectra show splitting because the induced magnetization and the resulting internal magnetic fields on the $^{1}$H nuclei become pronounced owing to the appearance of localized spins.
The magnetization, and consequently the splitting width, increase according to Curie's law with decreasing temperature.
Below 7 K, the spectrum becomes broad due to antiferromagnetic long-range ordering.
These results are consistent with a previous $^{1}$H-NMR study~\cite{Hiraki1995}.

Under non-equilibrium conditions with an applied current of 2.0 mA (Fig.~S1(b)), when the ambient temperature is above the transition temperature (79 K), the spectra show single peaks, identical to those under equilibrium conditions.
By contrast, when the ambient temperature is below the transition temperature and the intermediate resistance state is realized, the spectra deviate from those under equilibrium conditions.
In this region, the spectra show broadening owing to the appearance of the insulating phase.
However, spectral splitting is ambiguous, indicating that the local sample temperature in the insulating phase is somewhat distributed.
Even when the ambient temperature is decreased to absolute zero, almost the entire spectral range remains within $-$30 kHz to 60 kHz, indicating that the local sample temperature remains above $\sim$50 K almost everywhere, even in the region with the lowest local temperature.

To investigate the distribution of the local sample temperature, we analyzed the spectrum of the intermediate resistance state at 38.5 K (Fig.~S2(b)).
We reconstructed the spectrum by summing the spectra of the insulating phase under equilibrium conditions at 75.5 K, 65.7 K, and 56.1 K (Fig.~S2(a)).
Overall, the spectrum of the intermediate resistance state at 38.5 K closely resembles that of the insulating phase under equilibrium conditions at 75.5 K, indicating that the local temperature over most of the sample in the intermediate resistance state is nearly fixed at the transition temperature.
However, upon closer inspection, the spectral width of the intermediate resistance state at 38.5 K (75 kHz) is somewhat broader than that of the insulating state at 75.5 K (55 kHz), indicating the presence of slightly lower-temperature regions than 75.5 K.
Thus, we incorporated the broader spectra at 65.7 K and 56.1 K into the spectrum at 75.5 K to reconstruct the spectral shape of the intermediate resistance state at 38.5 K (Note that the volume fraction of the metallic phase at 38.5 K is $\sim$6\%, as shown in Fig. 8 of the main text; therefore, its contribution to the spectrum is negligible.)
The summed spectrum, created by adding the normalized spectra at 75.5 K, 65.7 K, and 56.1 K with an integrated-intensity ratio of 2 : 1 : 1 (corresponding to a volume fraction ratio of 2.69 : 1.17 : 1), successfully reproduces the spectrum of the intermediate resistance state at 38.5 K, as shown in Fig.~S2(b).
This result indicates that the local temperature over most of the sample is fixed around the transition temperature, but the local-temperature distribution extends slightly toward lower temperatures across the sample. This is likely the reason for the observation that the averaged sample temperature is slightly reduced from the transition temperature when the ambient temperature decreases, as shown in Fig. 6 of the main text.

\begin{figure}[bt]
\centering
\includegraphics[width=\hsize]{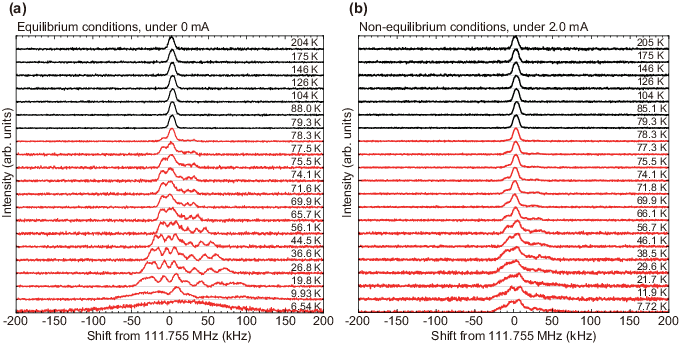}
\caption{
Ambient temperature dependence of the $^{1}$H-NMR spectra of ($d$7-DMe-DCNQI)$_{2}$Cu. The black lines represent the spectra above the transition temperature of 79 K, while the red lines represent those below the transition temperature. (a) Spectra under equilibrium conditions without an applied current. (b) Spectra under non-equilibrium conditions with an applied current of 2.0 mA.
}
\end{figure}

\begin{figure}[bt]
\centering
\includegraphics[width=\hsize]{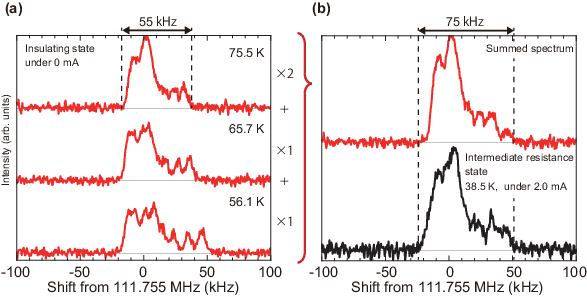}
\caption{
Reconstructed $^{1}$H-NMR spectrum of the intermediate resistance state. (a) $^{1}$H-NMR spectra of the insulating phase under equilibrium conditions without an applied current at 75.5 K, 65.7 K, and 56.1 K. The integrated intensities are normalized to be the same. (b) Reconstructed spectrum (red line), obtained by summing the three spectra in (a) with an integrated-intensity ratio of 2 : 1 : 1 (corresponding to a volume fraction ratio of 2.69 : 1.17 : 1), and the observed spectrum (black line) of the intermediate resistance state under 2.0 mA at 38.5 K. The reconstructed spectrum well reproduces the observed spectrum.
}
\end{figure}


\begin{thebibliography}{65}%
\makeatletter
\providecommand \@ifxundefined [1]{%
 \@ifx{#1\undefined}
}%
\providecommand \@ifnum [1]{%
 \ifnum #1\expandafter \@firstoftwo
 \else \expandafter \@secondoftwo
 \fi
}%
\providecommand \@ifx [1]{%
 \ifx #1\expandafter \@firstoftwo
 \else \expandafter \@secondoftwo
 \fi
}%
\providecommand \natexlab [1]{#1}%
\providecommand \enquote  [1]{``#1''}%
\providecommand \bibnamefont  [1]{#1}%
\providecommand \bibfnamefont [1]{#1}%
\providecommand \citenamefont [1]{#1}%
\providecommand \href@noop [0]{\@secondoftwo}%
\providecommand \href [0]{\begingroup \@sanitize@url \@href}%
\providecommand \@href[1]{\@@startlink{#1}\@@href}%
\providecommand \@@href[1]{\endgroup#1\@@endlink}%
\providecommand \@sanitize@url [0]{\catcode `\\12\catcode `\$12\catcode
  `\&12\catcode `\#12\catcode `\^12\catcode `\_12\catcode `\%12\relax}%
\providecommand \@@startlink[1]{}%
\providecommand \@@endlink[0]{}%
\providecommand \url  [0]{\begingroup\@sanitize@url \@url }%
\providecommand \@url [1]{\endgroup\@href {#1}{\urlprefix }}%
\providecommand \urlprefix  [0]{URL }%
\providecommand \Eprint [0]{\href }%
\providecommand \doibase [0]{https://doi.org/}%
\providecommand \selectlanguage [0]{\@gobble}%
\providecommand \bibinfo  [0]{\@secondoftwo}%
\providecommand \bibfield  [0]{\@secondoftwo}%
\providecommand \translation [1]{[#1]}%
\providecommand \BibitemOpen [0]{}%
\providecommand \bibitemStop [0]{}%
\providecommand \bibitemNoStop [0]{.\EOS\space}%
\providecommand \EOS [0]{\spacefactor3000\relax}%
\providecommand \BibitemShut  [1]{\csname bibitem#1\endcsname}%
\let\auto@bib@innerbib\@empty
\bibitem [{\citenamefont {Imada}\ \emph {et~al.}(1998)\citenamefont {Imada},
  \citenamefont {Fujimori},\ and\ \citenamefont {Tokura}}]{Imada1998}%
  \BibitemOpen
  \bibfield  {author} {\bibinfo {author} {\bibfnamefont {M.}~\bibnamefont
  {Imada}}, \bibinfo {author} {\bibfnamefont {A.}~\bibnamefont {Fujimori}},\
  and\ \bibinfo {author} {\bibfnamefont {Y.}~\bibnamefont {Tokura}},\
  }\bibfield  {title} {\bibinfo {title} {Metal-insulator transitions},\ }\href
  {https://doi.org/10.1103/RevModPhys.70.1039} {\bibfield  {journal} {\bibinfo
  {journal} {Reviews of Modern Physics}\ }\textbf {\bibinfo {volume} {70}},\
  \bibinfo {pages} {1039} (\bibinfo {year} {1998})}\BibitemShut {NoStop}%
\bibitem [{\citenamefont {Son}\ \emph {et~al.}(2011)\citenamefont {Son},
  \citenamefont {Lee}, \citenamefont {Park}, \citenamefont {Shin},
  \citenamefont {Choi}, \citenamefont {Jung}, \citenamefont {Lee},
  \citenamefont {Kim}, \citenamefont {Park},\ and\ \citenamefont
  {Hwang}}]{Son2011}%
  \BibitemOpen
  \bibfield  {author} {\bibinfo {author} {\bibfnamefont {M.}~\bibnamefont
  {Son}}, \bibinfo {author} {\bibfnamefont {J.}~\bibnamefont {Lee}}, \bibinfo
  {author} {\bibfnamefont {J.}~\bibnamefont {Park}}, \bibinfo {author}
  {\bibfnamefont {J.}~\bibnamefont {Shin}}, \bibinfo {author} {\bibfnamefont
  {G.}~\bibnamefont {Choi}}, \bibinfo {author} {\bibfnamefont {S.}~\bibnamefont
  {Jung}}, \bibinfo {author} {\bibfnamefont {W.}~\bibnamefont {Lee}}, \bibinfo
  {author} {\bibfnamefont {S.}~\bibnamefont {Kim}}, \bibinfo {author}
  {\bibfnamefont {S.}~\bibnamefont {Park}},\ and\ \bibinfo {author}
  {\bibfnamefont {H.}~\bibnamefont {Hwang}},\ }\bibfield  {title} {\bibinfo
  {title} {Excellent selector characteristics of nanoscale {VO}$_{2}$ for
  high-density bipolar {R}e{RAM} applications},\ }\href
  {https://doi.org/10.1109/LED.2011.2163697} {\bibfield  {journal} {\bibinfo
  {journal} {IEEE Electron Device Letters}\ }\textbf {\bibinfo {volume} {32}},\
  \bibinfo {pages} {1579} (\bibinfo {year} {2011})}\BibitemShut {NoStop}%
\bibitem [{\citenamefont {Chang}\ \emph {et~al.}(2011)\citenamefont {Chang},
  \citenamefont {Lee}, \citenamefont {Jeon}, \citenamefont {Park},
  \citenamefont {Kim}, \citenamefont {Yang}, \citenamefont {Chae},
  \citenamefont {Yoo}, \citenamefont {Kang}, \citenamefont {Lee},\ and\
  \citenamefont {Noh}}]{Chang2011}%
  \BibitemOpen
  \bibfield  {author} {\bibinfo {author} {\bibfnamefont {S.~H.}\ \bibnamefont
  {Chang}}, \bibinfo {author} {\bibfnamefont {S.~B.}\ \bibnamefont {Lee}},
  \bibinfo {author} {\bibfnamefont {D.~Y.}\ \bibnamefont {Jeon}}, \bibinfo
  {author} {\bibfnamefont {S.~J.}\ \bibnamefont {Park}}, \bibinfo {author}
  {\bibfnamefont {G.~T.}\ \bibnamefont {Kim}}, \bibinfo {author} {\bibfnamefont
  {S.~M.}\ \bibnamefont {Yang}}, \bibinfo {author} {\bibfnamefont {S.~C.}\
  \bibnamefont {Chae}}, \bibinfo {author} {\bibfnamefont {H.~K.}\ \bibnamefont
  {Yoo}}, \bibinfo {author} {\bibfnamefont {B.~S.}\ \bibnamefont {Kang}},
  \bibinfo {author} {\bibfnamefont {M.~J.}\ \bibnamefont {Lee}},\ and\ \bibinfo
  {author} {\bibfnamefont {T.~W.}\ \bibnamefont {Noh}},\ }\bibfield  {title}
  {\bibinfo {title} {Oxide double-layer nanocrossbar for ultrahigh-density
  bipolar resistive memory},\ }\href {https://doi.org/10.1002/adma.201102395}
  {\bibfield  {journal} {\bibinfo  {journal} {Advanced Materials}\ }\textbf
  {\bibinfo {volume} {23}},\ \bibinfo {pages} {4063} (\bibinfo {year}
  {2011})}\BibitemShut {NoStop}%
\bibitem [{\citenamefont {Pellegrino}\ \emph {et~al.}(2012)\citenamefont
  {Pellegrino}, \citenamefont {Manca}, \citenamefont {Kanki}, \citenamefont
  {Tanaka}, \citenamefont {Biasotti}, \citenamefont {Bellingeri}, \citenamefont
  {Siri},\ and\ \citenamefont {Marr\'{e}}}]{Pellegrino2012}%
  \BibitemOpen
  \bibfield  {author} {\bibinfo {author} {\bibfnamefont {L.}~\bibnamefont
  {Pellegrino}}, \bibinfo {author} {\bibfnamefont {N.}~\bibnamefont {Manca}},
  \bibinfo {author} {\bibfnamefont {T.}~\bibnamefont {Kanki}}, \bibinfo
  {author} {\bibfnamefont {H.}~\bibnamefont {Tanaka}}, \bibinfo {author}
  {\bibfnamefont {M.}~\bibnamefont {Biasotti}}, \bibinfo {author}
  {\bibfnamefont {E.}~\bibnamefont {Bellingeri}}, \bibinfo {author}
  {\bibfnamefont {A.~S.}\ \bibnamefont {Siri}},\ and\ \bibinfo {author}
  {\bibfnamefont {D.}~\bibnamefont {Marr\'{e}}},\ }\bibfield  {title} {\bibinfo
  {title} {Multistate memory devices based on free-standing
  {VO}$_{2}$/{TiO}$_{2}$ microstructures driven by joule self-heating},\ }\href
  {https://doi.org/10.1002/adma.201104669} {\bibfield  {journal} {\bibinfo
  {journal} {Advanced Materials}\ }\textbf {\bibinfo {volume} {24}},\ \bibinfo
  {pages} {2929} (\bibinfo {year} {2012})}\BibitemShut {NoStop}%
\bibitem [{\citenamefont {Bae}\ \emph {et~al.}(2013)\citenamefont {Bae},
  \citenamefont {Lee}, \citenamefont {Koo}, \citenamefont {Lin}, \citenamefont
  {Jo}, \citenamefont {Park},\ and\ \citenamefont {Wang}}]{Bae2013}%
  \BibitemOpen
  \bibfield  {author} {\bibinfo {author} {\bibfnamefont {S.~H.}\ \bibnamefont
  {Bae}}, \bibinfo {author} {\bibfnamefont {S.}~\bibnamefont {Lee}}, \bibinfo
  {author} {\bibfnamefont {H.}~\bibnamefont {Koo}}, \bibinfo {author}
  {\bibfnamefont {L.}~\bibnamefont {Lin}}, \bibinfo {author} {\bibfnamefont
  {B.~H.}\ \bibnamefont {Jo}}, \bibinfo {author} {\bibfnamefont
  {C.}~\bibnamefont {Park}},\ and\ \bibinfo {author} {\bibfnamefont {Z.~L.}\
  \bibnamefont {Wang}},\ }\bibfield  {title} {\bibinfo {title} {The memristive
  properties of a single {VO}$_{2}$ nanowire with switching controlled by
  self-heating},\ }\href {https://doi.org/10.1002/adma.201302511} {\bibfield
  {journal} {\bibinfo  {journal} {Advanced Materials}\ }\textbf {\bibinfo
  {volume} {25}},\ \bibinfo {pages} {5098} (\bibinfo {year}
  {2013})}\BibitemShut {NoStop}%
\bibitem [{\citenamefont {Janod}\ \emph {et~al.}(2015)\citenamefont {Janod},
  \citenamefont {Tranchant}, \citenamefont {Corraze}, \citenamefont
  {Querr\'{e}}, \citenamefont {Stoliar}, \citenamefont {Rozenberg},
  \citenamefont {Cren}, \citenamefont {Roditchev}, \citenamefont {Phuoc},
  \citenamefont {Besland},\ and\ \citenamefont {Cario}}]{Janod2015}%
  \BibitemOpen
  \bibfield  {author} {\bibinfo {author} {\bibfnamefont {E.}~\bibnamefont
  {Janod}}, \bibinfo {author} {\bibfnamefont {J.}~\bibnamefont {Tranchant}},
  \bibinfo {author} {\bibfnamefont {B.}~\bibnamefont {Corraze}}, \bibinfo
  {author} {\bibfnamefont {M.}~\bibnamefont {Querr\'{e}}}, \bibinfo {author}
  {\bibfnamefont {P.}~\bibnamefont {Stoliar}}, \bibinfo {author} {\bibfnamefont
  {M.}~\bibnamefont {Rozenberg}}, \bibinfo {author} {\bibfnamefont
  {T.}~\bibnamefont {Cren}}, \bibinfo {author} {\bibfnamefont {D.}~\bibnamefont
  {Roditchev}}, \bibinfo {author} {\bibfnamefont {V.~T.}\ \bibnamefont
  {Phuoc}}, \bibinfo {author} {\bibfnamefont {M.~P.}\ \bibnamefont {Besland}},\
  and\ \bibinfo {author} {\bibfnamefont {L.}~\bibnamefont {Cario}},\ }\bibfield
   {title} {\bibinfo {title} {Resistive {S}witching in {M}ott {I}nsulators and
  {C}orrelated {S}ystems},\ }\href {https://doi.org/10.1002/adfm.201500823}
  {\bibfield  {journal} {\bibinfo  {journal} {Advanced Functional Materials}\
  }\textbf {\bibinfo {volume} {25}},\ \bibinfo {pages} {6287} (\bibinfo {year}
  {2015})}\BibitemShut {NoStop}%
\bibitem [{\citenamefont {Coy}\ \emph {et~al.}(2010)\citenamefont {Coy},
  \citenamefont {Cabrera}, \citenamefont {Sep\'{u}lveda},\ and\ \citenamefont
  {Fern\'{a}ndez}}]{Coy2010}%
  \BibitemOpen
  \bibfield  {author} {\bibinfo {author} {\bibfnamefont {H.}~\bibnamefont
  {Coy}}, \bibinfo {author} {\bibfnamefont {R.}~\bibnamefont {Cabrera}},
  \bibinfo {author} {\bibfnamefont {N.}~\bibnamefont {Sep\'{u}lveda}},\ and\
  \bibinfo {author} {\bibfnamefont {F.~E.}\ \bibnamefont {Fern\'{a}ndez}},\
  }\bibfield  {title} {\bibinfo {title} {Optoelectronic and all-optical
  multiple memory states in vanadium dioxide},\ }\href
  {https://doi.org/10.1063/1.3518508} {\bibfield  {journal} {\bibinfo
  {journal} {Journal of Applied Physics}\ }\textbf {\bibinfo {volume} {108}},\
  \bibinfo {pages} {113115} (\bibinfo {year} {2010})}\BibitemShut {NoStop}%
\bibitem [{\citenamefont {Liu}\ \emph {et~al.}(2012)\citenamefont {Liu},
  \citenamefont {Hwang}, \citenamefont {Tao}, \citenamefont {Strikwerda},
  \citenamefont {Fan}, \citenamefont {Keiser}, \citenamefont {Sternbach},
  \citenamefont {West}, \citenamefont {Kittiwatanakul}, \citenamefont {Lu},
  \citenamefont {Wolf}, \citenamefont {Omenetto}, \citenamefont {Zhang},
  \citenamefont {Nelson},\ and\ \citenamefont {Averitt}}]{Liu2012}%
  \BibitemOpen
  \bibfield  {author} {\bibinfo {author} {\bibfnamefont {M.}~\bibnamefont
  {Liu}}, \bibinfo {author} {\bibfnamefont {H.~Y.}\ \bibnamefont {Hwang}},
  \bibinfo {author} {\bibfnamefont {H.}~\bibnamefont {Tao}}, \bibinfo {author}
  {\bibfnamefont {A.~C.}\ \bibnamefont {Strikwerda}}, \bibinfo {author}
  {\bibfnamefont {K.}~\bibnamefont {Fan}}, \bibinfo {author} {\bibfnamefont
  {G.~R.}\ \bibnamefont {Keiser}}, \bibinfo {author} {\bibfnamefont {A.~J.}\
  \bibnamefont {Sternbach}}, \bibinfo {author} {\bibfnamefont {K.~G.}\
  \bibnamefont {West}}, \bibinfo {author} {\bibfnamefont {S.}~\bibnamefont
  {Kittiwatanakul}}, \bibinfo {author} {\bibfnamefont {J.}~\bibnamefont {Lu}},
  \bibinfo {author} {\bibfnamefont {S.~A.}\ \bibnamefont {Wolf}}, \bibinfo
  {author} {\bibfnamefont {F.~G.}\ \bibnamefont {Omenetto}}, \bibinfo {author}
  {\bibfnamefont {X.}~\bibnamefont {Zhang}}, \bibinfo {author} {\bibfnamefont
  {K.~A.}\ \bibnamefont {Nelson}},\ and\ \bibinfo {author} {\bibfnamefont
  {R.~D.}\ \bibnamefont {Averitt}},\ }\bibfield  {title} {\bibinfo {title}
  {Terahertz-field-induced insulator-to-metal transition in vanadium dioxide
  metamaterial},\ }\href {https://doi.org/10.1038/nature11231} {\bibfield
  {journal} {\bibinfo  {journal} {Nature}\ }\textbf {\bibinfo {volume} {487}},\
  \bibinfo {pages} {345} (\bibinfo {year} {2012})}\BibitemShut {NoStop}%
\bibitem [{\citenamefont {Markov}\ \emph {et~al.}(2015)\citenamefont {Markov},
  \citenamefont {Marvel}, \citenamefont {Conley}, \citenamefont {Miller},
  \citenamefont {Haglund},\ and\ \citenamefont {Weiss}}]{Markov2015}%
  \BibitemOpen
  \bibfield  {author} {\bibinfo {author} {\bibfnamefont {P.}~\bibnamefont
  {Markov}}, \bibinfo {author} {\bibfnamefont {R.~E.}\ \bibnamefont {Marvel}},
  \bibinfo {author} {\bibfnamefont {H.~J.}\ \bibnamefont {Conley}}, \bibinfo
  {author} {\bibfnamefont {K.~J.}\ \bibnamefont {Miller}}, \bibinfo {author}
  {\bibfnamefont {R.~F.}\ \bibnamefont {Haglund}},\ and\ \bibinfo {author}
  {\bibfnamefont {S.~M.}\ \bibnamefont {Weiss}},\ }\bibfield  {title} {\bibinfo
  {title} {Optically {M}onitored {E}lectrical {S}witching in {VO}$_{2}$},\
  }\href {https://doi.org/10.1021/acsphotonics.5b00244} {\bibfield  {journal}
  {\bibinfo  {journal} {ACS Photonics}\ }\textbf {\bibinfo {volume} {2}},\
  \bibinfo {pages} {1175} (\bibinfo {year} {2015})}\BibitemShut {NoStop}%
\bibitem [{\citenamefont {Fan}\ \emph {et~al.}(2016)\citenamefont {Fan},
  \citenamefont {Chen}, \citenamefont {Liu}, \citenamefont {Chen},
  \citenamefont {Zhu}, \citenamefont {Meng}, \citenamefont {Wang},
  \citenamefont {Zhang}, \citenamefont {Ren},\ and\ \citenamefont
  {Zou}}]{Fan2016}%
  \BibitemOpen
  \bibfield  {author} {\bibinfo {author} {\bibfnamefont {L.}~\bibnamefont
  {Fan}}, \bibinfo {author} {\bibfnamefont {Y.}~\bibnamefont {Chen}}, \bibinfo
  {author} {\bibfnamefont {Q.}~\bibnamefont {Liu}}, \bibinfo {author}
  {\bibfnamefont {S.}~\bibnamefont {Chen}}, \bibinfo {author} {\bibfnamefont
  {L.}~\bibnamefont {Zhu}}, \bibinfo {author} {\bibfnamefont {Q.}~\bibnamefont
  {Meng}}, \bibinfo {author} {\bibfnamefont {B.}~\bibnamefont {Wang}}, \bibinfo
  {author} {\bibfnamefont {Q.}~\bibnamefont {Zhang}}, \bibinfo {author}
  {\bibfnamefont {H.}~\bibnamefont {Ren}},\ and\ \bibinfo {author}
  {\bibfnamefont {C.}~\bibnamefont {Zou}},\ }\bibfield  {title} {\bibinfo
  {title} {Infrared {R}esponse and {O}ptoelectronic {M}emory {D}evice
  {F}abrication {B}ased on {E}pitaxial {VO}$_{2}$ {F}ilm},\ }\href
  {https://doi.org/10.1021/acsami.6b12831} {\bibfield  {journal} {\bibinfo
  {journal} {ACS Applied Materials and Interfaces}\ }\textbf {\bibinfo {volume}
  {8}},\ \bibinfo {pages} {32971} (\bibinfo {year} {2016})}\BibitemShut
  {NoStop}%
\bibitem [{\citenamefont {Lee}\ \emph {et~al.}(2017)\citenamefont {Lee},
  \citenamefont {Lee}, \citenamefont {Song}, \citenamefont {Xue}, \citenamefont
  {Choi}, \citenamefont {Ma}, \citenamefont {Podkaminer}, \citenamefont {Liu},
  \citenamefont {Liu}, \citenamefont {Chung}, \citenamefont {Fan},
  \citenamefont {Cho}, \citenamefont {Zhou}, \citenamefont {Lee}, \citenamefont
  {Chen}, \citenamefont {Oh}, \citenamefont {Ma},\ and\ \citenamefont
  {Eom}}]{Lee2017}%
  \BibitemOpen
  \bibfield  {author} {\bibinfo {author} {\bibfnamefont {D.}~\bibnamefont
  {Lee}}, \bibinfo {author} {\bibfnamefont {J.}~\bibnamefont {Lee}}, \bibinfo
  {author} {\bibfnamefont {K.}~\bibnamefont {Song}}, \bibinfo {author}
  {\bibfnamefont {F.}~\bibnamefont {Xue}}, \bibinfo {author} {\bibfnamefont
  {S.~Y.}\ \bibnamefont {Choi}}, \bibinfo {author} {\bibfnamefont
  {Y.}~\bibnamefont {Ma}}, \bibinfo {author} {\bibfnamefont {J.}~\bibnamefont
  {Podkaminer}}, \bibinfo {author} {\bibfnamefont {D.}~\bibnamefont {Liu}},
  \bibinfo {author} {\bibfnamefont {S.~C.}\ \bibnamefont {Liu}}, \bibinfo
  {author} {\bibfnamefont {B.}~\bibnamefont {Chung}}, \bibinfo {author}
  {\bibfnamefont {W.}~\bibnamefont {Fan}}, \bibinfo {author} {\bibfnamefont
  {S.~J.}\ \bibnamefont {Cho}}, \bibinfo {author} {\bibfnamefont
  {W.}~\bibnamefont {Zhou}}, \bibinfo {author} {\bibfnamefont {J.}~\bibnamefont
  {Lee}}, \bibinfo {author} {\bibfnamefont {L.~Q.}\ \bibnamefont {Chen}},
  \bibinfo {author} {\bibfnamefont {S.~H.}\ \bibnamefont {Oh}}, \bibinfo
  {author} {\bibfnamefont {Z.}~\bibnamefont {Ma}},\ and\ \bibinfo {author}
  {\bibfnamefont {C.~B.}\ \bibnamefont {Eom}},\ }\bibfield  {title} {\bibinfo
  {title} {Sharpened {VO}$_{2}$ {P}hase {T}ransition via {C}ontrolled {R}elease
  of {E}pitaxial {S}train},\ }\href
  {https://doi.org/10.1021/acs.nanolett.7b02482} {\bibfield  {journal}
  {\bibinfo  {journal} {Nano Letters}\ }\textbf {\bibinfo {volume} {17}},\
  \bibinfo {pages} {5614} (\bibinfo {year} {2017})}\BibitemShut {NoStop}%
\bibitem [{\citenamefont {Miller}\ \emph {et~al.}(2017)\citenamefont {Miller},
  \citenamefont {Hallman}, \citenamefont {Haglund},\ and\ \citenamefont
  {Weiss}}]{Miller2017}%
  \BibitemOpen
  \bibfield  {author} {\bibinfo {author} {\bibfnamefont {K.~J.}\ \bibnamefont
  {Miller}}, \bibinfo {author} {\bibfnamefont {K.~A.}\ \bibnamefont {Hallman}},
  \bibinfo {author} {\bibfnamefont {R.~F.}\ \bibnamefont {Haglund}},\ and\
  \bibinfo {author} {\bibfnamefont {S.~M.}\ \bibnamefont {Weiss}},\ }\bibfield
  {title} {\bibinfo {title} {Silicon waveguide optical switch with embedded
  phase change material},\ }\href {https://doi.org/10.1364/oe.25.026527}
  {\bibfield  {journal} {\bibinfo  {journal} {Optics Express}\ }\textbf
  {\bibinfo {volume} {25}},\ \bibinfo {pages} {26527} (\bibinfo {year}
  {2017})}\BibitemShut {NoStop}%
\bibitem [{\citenamefont {Butakov}\ \emph
  {et~al.}(2018{\natexlab{a}})\citenamefont {Butakov}, \citenamefont {Knight},
  \citenamefont {Lewi}, \citenamefont {Iyer}, \citenamefont {Higgs},
  \citenamefont {Chorsi}, \citenamefont {Trastoy}, \citenamefont {Granda},
  \citenamefont {Valmianski}, \citenamefont {Urban}, \citenamefont {Kalcheim},
  \citenamefont {Wang}, \citenamefont {Hon}, \citenamefont {Schuller},\ and\
  \citenamefont {Schuller}}]{Butakov2018}%
  \BibitemOpen
  \bibfield  {author} {\bibinfo {author} {\bibfnamefont {N.~A.}\ \bibnamefont
  {Butakov}}, \bibinfo {author} {\bibfnamefont {M.~W.}\ \bibnamefont {Knight}},
  \bibinfo {author} {\bibfnamefont {T.}~\bibnamefont {Lewi}}, \bibinfo {author}
  {\bibfnamefont {P.~P.}\ \bibnamefont {Iyer}}, \bibinfo {author}
  {\bibfnamefont {D.}~\bibnamefont {Higgs}}, \bibinfo {author} {\bibfnamefont
  {H.~T.}\ \bibnamefont {Chorsi}}, \bibinfo {author} {\bibfnamefont
  {J.}~\bibnamefont {Trastoy}}, \bibinfo {author} {\bibfnamefont {J.~D.~V.}\
  \bibnamefont {Granda}}, \bibinfo {author} {\bibfnamefont {I.}~\bibnamefont
  {Valmianski}}, \bibinfo {author} {\bibfnamefont {C.}~\bibnamefont {Urban}},
  \bibinfo {author} {\bibfnamefont {Y.}~\bibnamefont {Kalcheim}}, \bibinfo
  {author} {\bibfnamefont {P.~Y.}\ \bibnamefont {Wang}}, \bibinfo {author}
  {\bibfnamefont {P.~W.}\ \bibnamefont {Hon}}, \bibinfo {author} {\bibfnamefont
  {I.~K.}\ \bibnamefont {Schuller}},\ and\ \bibinfo {author} {\bibfnamefont
  {J.~A.}\ \bibnamefont {Schuller}},\ }\bibfield  {title} {\bibinfo {title}
  {Broadband {E}lectrically {T}unable {D}ielectric {R}esonators {U}sing
  {M}etal-{I}nsulator {T}ransitions},\ }\href
  {https://doi.org/10.1021/acsphotonics.8b00699} {\bibfield  {journal}
  {\bibinfo  {journal} {ACS Photonics}\ }\textbf {\bibinfo {volume} {5}},\
  \bibinfo {pages} {4056} (\bibinfo {year} {2018}{\natexlab{a}})}\BibitemShut
  {NoStop}%
\bibitem [{\citenamefont {S\'{a}nchez}\ \emph {et~al.}(2018)\citenamefont
  {S\'{a}nchez}, \citenamefont {Olivares}, \citenamefont {Parra}, \citenamefont
  {Menghini}, \citenamefont {Homm}, \citenamefont {Locquet},\ and\
  \citenamefont {Sanchis}}]{Sanchez2018}%
  \BibitemOpen
  \bibfield  {author} {\bibinfo {author} {\bibfnamefont {L.~D.}\ \bibnamefont
  {S\'{a}nchez}}, \bibinfo {author} {\bibfnamefont {I.}~\bibnamefont
  {Olivares}}, \bibinfo {author} {\bibfnamefont {J.}~\bibnamefont {Parra}},
  \bibinfo {author} {\bibfnamefont {M.}~\bibnamefont {Menghini}}, \bibinfo
  {author} {\bibfnamefont {P.}~\bibnamefont {Homm}}, \bibinfo {author}
  {\bibfnamefont {J.-P.}\ \bibnamefont {Locquet}},\ and\ \bibinfo {author}
  {\bibfnamefont {P.}~\bibnamefont {Sanchis}},\ }\bibfield  {title} {\bibinfo
  {title} {Experimental demonstration of a tunable transverse electric pass
  polarizer based on hybrid {VO}$_{2}$/silicon technology},\ }\href
  {https://doi.org/10.1364/ol.43.003650} {\bibfield  {journal} {\bibinfo
  {journal} {Optics Letters}\ }\textbf {\bibinfo {volume} {43}},\ \bibinfo
  {pages} {3650} (\bibinfo {year} {2018})}\BibitemShut {NoStop}%
\bibitem [{\citenamefont {Olivares}\ \emph {et~al.}(2018)\citenamefont
  {Olivares}, \citenamefont {S\'{a}nchez}, \citenamefont {Parra}, \citenamefont
  {Larrea}, \citenamefont {Griol}, \citenamefont {Menghini}, \citenamefont
  {Homm}, \citenamefont {Jang}, \citenamefont {van Bilzen}, \citenamefont
  {Seo}, \citenamefont {Locquet},\ and\ \citenamefont
  {Sanchis}}]{Olivares2018}%
  \BibitemOpen
  \bibfield  {author} {\bibinfo {author} {\bibfnamefont {I.}~\bibnamefont
  {Olivares}}, \bibinfo {author} {\bibfnamefont {L.}~\bibnamefont
  {S\'{a}nchez}}, \bibinfo {author} {\bibfnamefont {J.}~\bibnamefont {Parra}},
  \bibinfo {author} {\bibfnamefont {R.}~\bibnamefont {Larrea}}, \bibinfo
  {author} {\bibfnamefont {A.}~\bibnamefont {Griol}}, \bibinfo {author}
  {\bibfnamefont {M.}~\bibnamefont {Menghini}}, \bibinfo {author}
  {\bibfnamefont {P.}~\bibnamefont {Homm}}, \bibinfo {author} {\bibfnamefont
  {L.-W.}\ \bibnamefont {Jang}}, \bibinfo {author} {\bibfnamefont
  {B.}~\bibnamefont {van Bilzen}}, \bibinfo {author} {\bibfnamefont {J.~W.}\
  \bibnamefont {Seo}}, \bibinfo {author} {\bibfnamefont {J.-P.}\ \bibnamefont
  {Locquet}},\ and\ \bibinfo {author} {\bibfnamefont {P.}~\bibnamefont
  {Sanchis}},\ }\bibfield  {title} {\bibinfo {title} {Optical switching in
  hybrid {VO}$_{2}$/{S}i waveguides thermally triggered by lateral
  microheaters},\ }\href {https://doi.org/10.1364/oe.26.012387} {\bibfield
  {journal} {\bibinfo  {journal} {Optics Express}\ }\textbf {\bibinfo {volume}
  {26}},\ \bibinfo {pages} {12387} (\bibinfo {year} {2018})}\BibitemShut
  {NoStop}%
\bibitem [{\citenamefont {Butakov}\ \emph
  {et~al.}(2018{\natexlab{b}})\citenamefont {Butakov}, \citenamefont
  {Valmianski}, \citenamefont {Lewi}, \citenamefont {Urban}, \citenamefont
  {Ren}, \citenamefont {Mikhailovsky}, \citenamefont {Wilson}, \citenamefont
  {Schuller},\ and\ \citenamefont {Schuller}}]{Butakov2018_2}%
  \BibitemOpen
  \bibfield  {author} {\bibinfo {author} {\bibfnamefont {N.~A.}\ \bibnamefont
  {Butakov}}, \bibinfo {author} {\bibfnamefont {I.}~\bibnamefont {Valmianski}},
  \bibinfo {author} {\bibfnamefont {T.}~\bibnamefont {Lewi}}, \bibinfo {author}
  {\bibfnamefont {C.}~\bibnamefont {Urban}}, \bibinfo {author} {\bibfnamefont
  {Z.}~\bibnamefont {Ren}}, \bibinfo {author} {\bibfnamefont {A.~A.}\
  \bibnamefont {Mikhailovsky}}, \bibinfo {author} {\bibfnamefont {S.~D.}\
  \bibnamefont {Wilson}}, \bibinfo {author} {\bibfnamefont {I.~K.}\
  \bibnamefont {Schuller}},\ and\ \bibinfo {author} {\bibfnamefont {J.~A.}\
  \bibnamefont {Schuller}},\ }\bibfield  {title} {\bibinfo {title} {Switchable
  {P}lasmonic-{D}ielectric {R}esonators with {M}etal-{I}nsulator
  {T}ransitions},\ }\href {https://doi.org/10.1021/acsphotonics.7b00334}
  {\bibfield  {journal} {\bibinfo  {journal} {ACS Photonics}\ }\textbf
  {\bibinfo {volume} {5}},\ \bibinfo {pages} {371} (\bibinfo {year}
  {2018}{\natexlab{b}})}\BibitemShut {NoStop}%
\bibitem [{\citenamefont {Coll}\ \emph {et~al.}(2019)\citenamefont {Coll},
  \citenamefont {Fontcuberta}, \citenamefont {Althammer}, \citenamefont
  {Bibes}, \citenamefont {Boschker}, \citenamefont {Calleja}, \citenamefont
  {Cheng}, \citenamefont {Cuoco}, \citenamefont {Dittmann}, \citenamefont
  {Dkhil}, \citenamefont {Baggari}, \citenamefont {Fanciulli}, \citenamefont
  {Fina}, \citenamefont {Fortunato}, \citenamefont {Frontera}, \citenamefont
  {Fujita}, \citenamefont {Garcia}, \citenamefont {Goennenwein}, \citenamefont
  {Granqvist}, \citenamefont {Grollier}, \citenamefont {Gross}, \citenamefont
  {Hagfeldt}, \citenamefont {Herranz}, \citenamefont {Hono}, \citenamefont
  {Houwman}, \citenamefont {Huijben}, \citenamefont {Kalaboukhov},
  \citenamefont {Keeble}, \citenamefont {Koster}, \citenamefont {Kourkoutis},
  \citenamefont {Levy}, \citenamefont {Lira-Cantu}, \citenamefont
  {MacManus-Driscoll}, \citenamefont {Mannhart}, \citenamefont {Martins},
  \citenamefont {Menzel}, \citenamefont {Mikolajick}, \citenamefont {Napari},
  \citenamefont {Nguyen}, \citenamefont {Niklasson}, \citenamefont {Paillard},
  \citenamefont {Panigrahi}, \citenamefont {Rijnders}, \citenamefont
  {S\'{a}nchez}, \citenamefont {Sanchis}, \citenamefont {Sanna}, \citenamefont
  {Schlom}, \citenamefont {Schroeder}, \citenamefont {Shen}, \citenamefont
  {Siemon}, \citenamefont {Spreitzer}, \citenamefont {Sukegawa}, \citenamefont
  {Tamayo}, \citenamefont {van~den Brink}, \citenamefont {Pryds},\ and\
  \citenamefont {Granozio}}]{Coll2019}%
  \BibitemOpen
  \bibfield  {author} {\bibinfo {author} {\bibfnamefont {M.}~\bibnamefont
  {Coll}}, \bibinfo {author} {\bibfnamefont {J.}~\bibnamefont {Fontcuberta}},
  \bibinfo {author} {\bibfnamefont {M.}~\bibnamefont {Althammer}}, \bibinfo
  {author} {\bibfnamefont {M.}~\bibnamefont {Bibes}}, \bibinfo {author}
  {\bibfnamefont {H.}~\bibnamefont {Boschker}}, \bibinfo {author}
  {\bibfnamefont {A.}~\bibnamefont {Calleja}}, \bibinfo {author} {\bibfnamefont
  {G.}~\bibnamefont {Cheng}}, \bibinfo {author} {\bibfnamefont
  {M.}~\bibnamefont {Cuoco}}, \bibinfo {author} {\bibfnamefont
  {R.}~\bibnamefont {Dittmann}}, \bibinfo {author} {\bibfnamefont
  {B.}~\bibnamefont {Dkhil}}, \bibinfo {author} {\bibfnamefont {I.~E.}\
  \bibnamefont {Baggari}}, \bibinfo {author} {\bibfnamefont {M.}~\bibnamefont
  {Fanciulli}}, \bibinfo {author} {\bibfnamefont {I.}~\bibnamefont {Fina}},
  \bibinfo {author} {\bibfnamefont {E.}~\bibnamefont {Fortunato}}, \bibinfo
  {author} {\bibfnamefont {C.}~\bibnamefont {Frontera}}, \bibinfo {author}
  {\bibfnamefont {S.}~\bibnamefont {Fujita}}, \bibinfo {author} {\bibfnamefont
  {V.}~\bibnamefont {Garcia}}, \bibinfo {author} {\bibfnamefont
  {S.}~\bibnamefont {Goennenwein}}, \bibinfo {author} {\bibfnamefont {C.-G.}\
  \bibnamefont {Granqvist}}, \bibinfo {author} {\bibfnamefont {J.}~\bibnamefont
  {Grollier}}, \bibinfo {author} {\bibfnamefont {R.}~\bibnamefont {Gross}},
  \bibinfo {author} {\bibfnamefont {A.}~\bibnamefont {Hagfeldt}}, \bibinfo
  {author} {\bibfnamefont {G.}~\bibnamefont {Herranz}}, \bibinfo {author}
  {\bibfnamefont {K.}~\bibnamefont {Hono}}, \bibinfo {author} {\bibfnamefont
  {E.}~\bibnamefont {Houwman}}, \bibinfo {author} {\bibfnamefont
  {M.}~\bibnamefont {Huijben}}, \bibinfo {author} {\bibfnamefont
  {A.}~\bibnamefont {Kalaboukhov}}, \bibinfo {author} {\bibfnamefont
  {D.}~\bibnamefont {Keeble}}, \bibinfo {author} {\bibfnamefont
  {G.}~\bibnamefont {Koster}}, \bibinfo {author} {\bibfnamefont
  {L.}~\bibnamefont {Kourkoutis}}, \bibinfo {author} {\bibfnamefont
  {J.}~\bibnamefont {Levy}}, \bibinfo {author} {\bibfnamefont {M.}~\bibnamefont
  {Lira-Cantu}}, \bibinfo {author} {\bibfnamefont {J.}~\bibnamefont
  {MacManus-Driscoll}}, \bibinfo {author} {\bibfnamefont {J.}~\bibnamefont
  {Mannhart}}, \bibinfo {author} {\bibfnamefont {R.}~\bibnamefont {Martins}},
  \bibinfo {author} {\bibfnamefont {S.}~\bibnamefont {Menzel}}, \bibinfo
  {author} {\bibfnamefont {T.}~\bibnamefont {Mikolajick}}, \bibinfo {author}
  {\bibfnamefont {M.}~\bibnamefont {Napari}}, \bibinfo {author} {\bibfnamefont
  {M.}~\bibnamefont {Nguyen}}, \bibinfo {author} {\bibfnamefont
  {G.}~\bibnamefont {Niklasson}}, \bibinfo {author} {\bibfnamefont
  {C.}~\bibnamefont {Paillard}}, \bibinfo {author} {\bibfnamefont
  {S.}~\bibnamefont {Panigrahi}}, \bibinfo {author} {\bibfnamefont
  {G.}~\bibnamefont {Rijnders}}, \bibinfo {author} {\bibfnamefont
  {F.}~\bibnamefont {S\'{a}nchez}}, \bibinfo {author} {\bibfnamefont
  {P.}~\bibnamefont {Sanchis}}, \bibinfo {author} {\bibfnamefont
  {S.}~\bibnamefont {Sanna}}, \bibinfo {author} {\bibfnamefont
  {D.}~\bibnamefont {Schlom}}, \bibinfo {author} {\bibfnamefont
  {U.}~\bibnamefont {Schroeder}}, \bibinfo {author} {\bibfnamefont
  {K.}~\bibnamefont {Shen}}, \bibinfo {author} {\bibfnamefont {A.}~\bibnamefont
  {Siemon}}, \bibinfo {author} {\bibfnamefont {M.}~\bibnamefont {Spreitzer}},
  \bibinfo {author} {\bibfnamefont {H.}~\bibnamefont {Sukegawa}}, \bibinfo
  {author} {\bibfnamefont {R.}~\bibnamefont {Tamayo}}, \bibinfo {author}
  {\bibfnamefont {J.}~\bibnamefont {van~den Brink}}, \bibinfo {author}
  {\bibfnamefont {N.}~\bibnamefont {Pryds}},\ and\ \bibinfo {author}
  {\bibfnamefont {F.~M.}\ \bibnamefont {Granozio}},\ }\bibfield  {title}
  {\bibinfo {title} {Towards {O}xide {E}lectronics: a {R}oadmap},\ }\href
  {https://doi.org/10.1016/j.apsusc.2019.03.312} {\bibfield  {journal}
  {\bibinfo  {journal} {Applied Surface Science}\ }\textbf {\bibinfo {volume}
  {482}},\ \bibinfo {pages} {1} (\bibinfo {year} {2019})}\BibitemShut {NoStop}%
\bibitem [{\citenamefont {Cueff}\ \emph {et~al.}(2020)\citenamefont {Cueff},
  \citenamefont {John}, \citenamefont {Zhang}, \citenamefont {Parra},
  \citenamefont {Sun}, \citenamefont {Orobtchouk}, \citenamefont {Ramanathan},\
  and\ \citenamefont {Sanchis}}]{Cueff2020}%
  \BibitemOpen
  \bibfield  {author} {\bibinfo {author} {\bibfnamefont {S.}~\bibnamefont
  {Cueff}}, \bibinfo {author} {\bibfnamefont {J.}~\bibnamefont {John}},
  \bibinfo {author} {\bibfnamefont {Z.}~\bibnamefont {Zhang}}, \bibinfo
  {author} {\bibfnamefont {J.}~\bibnamefont {Parra}}, \bibinfo {author}
  {\bibfnamefont {J.}~\bibnamefont {Sun}}, \bibinfo {author} {\bibfnamefont
  {R.}~\bibnamefont {Orobtchouk}}, \bibinfo {author} {\bibfnamefont
  {S.}~\bibnamefont {Ramanathan}},\ and\ \bibinfo {author} {\bibfnamefont
  {P.}~\bibnamefont {Sanchis}},\ }\bibfield  {title} {\bibinfo {title}
  {{VO}$_{2}$ nanophotonics},\ }\href {https://doi.org/10.1063/5.0028093}
  {\bibfield  {journal} {\bibinfo  {journal} {APL Photonics}\ }\textbf
  {\bibinfo {volume} {5}},\ \bibinfo {pages} {110901} (\bibinfo {year}
  {2020})}\BibitemShut {NoStop}%
\bibitem [{\citenamefont {Li}\ \emph {et~al.}(2022)\citenamefont {Li},
  \citenamefont {Xie}, \citenamefont {Zhong}, \citenamefont {Zhang},
  \citenamefont {Fu}, \citenamefont {Zhou}, \citenamefont {Li}, \citenamefont
  {Ni}, \citenamefont {Wang}, \citenamefont {jia Guo}, \citenamefont {He},
  \citenamefont {Wang}, \citenamefont {Yang}, \citenamefont {Jin},\ and\
  \citenamefont {Ge}}]{Li2022}%
  \BibitemOpen
  \bibfield  {author} {\bibinfo {author} {\bibfnamefont {G.}~\bibnamefont
  {Li}}, \bibinfo {author} {\bibfnamefont {D.}~\bibnamefont {Xie}}, \bibinfo
  {author} {\bibfnamefont {H.}~\bibnamefont {Zhong}}, \bibinfo {author}
  {\bibfnamefont {Z.}~\bibnamefont {Zhang}}, \bibinfo {author} {\bibfnamefont
  {X.}~\bibnamefont {Fu}}, \bibinfo {author} {\bibfnamefont {Q.}~\bibnamefont
  {Zhou}}, \bibinfo {author} {\bibfnamefont {Q.}~\bibnamefont {Li}}, \bibinfo
  {author} {\bibfnamefont {H.}~\bibnamefont {Ni}}, \bibinfo {author}
  {\bibfnamefont {J.}~\bibnamefont {Wang}}, \bibinfo {author} {\bibfnamefont
  {E.}~\bibnamefont {jia Guo}}, \bibinfo {author} {\bibfnamefont
  {M.}~\bibnamefont {He}}, \bibinfo {author} {\bibfnamefont {C.}~\bibnamefont
  {Wang}}, \bibinfo {author} {\bibfnamefont {G.}~\bibnamefont {Yang}}, \bibinfo
  {author} {\bibfnamefont {K.}~\bibnamefont {Jin}},\ and\ \bibinfo {author}
  {\bibfnamefont {C.}~\bibnamefont {Ge}},\ }\bibfield  {title} {\bibinfo
  {title} {Photo-induced non-volatile {VO}$_{2}$ phase transition for
  neuromorphic ultraviolet sensors},\ }\href
  {https://doi.org/10.1038/s41467-022-29456-5} {\bibfield  {journal} {\bibinfo
  {journal} {Nature Communications}\ }\textbf {\bibinfo {volume} {13}},\
  \bibinfo {pages} {1729} (\bibinfo {year} {2022})}\BibitemShut {NoStop}%
\bibitem [{\citenamefont {Pickett}\ \emph {et~al.}(2013)\citenamefont
  {Pickett}, \citenamefont {Medeiros-Ribeiro},\ and\ \citenamefont
  {Williams}}]{Pickett2013}%
  \BibitemOpen
  \bibfield  {author} {\bibinfo {author} {\bibfnamefont {M.~D.}\ \bibnamefont
  {Pickett}}, \bibinfo {author} {\bibfnamefont {G.}~\bibnamefont
  {Medeiros-Ribeiro}},\ and\ \bibinfo {author} {\bibfnamefont {R.~S.}\
  \bibnamefont {Williams}},\ }\bibfield  {title} {\bibinfo {title} {A scalable
  neuristor built with {M}ott memristors},\ }\href
  {https://doi.org/10.1038/nmat3510} {\bibfield  {journal} {\bibinfo  {journal}
  {Nature Materials}\ }\textbf {\bibinfo {volume} {12}},\ \bibinfo {pages}
  {114} (\bibinfo {year} {2013})}\BibitemShut {NoStop}%
\bibitem [{\citenamefont {Ignatov}\ \emph {et~al.}(2015)\citenamefont
  {Ignatov}, \citenamefont {Ziegler}, \citenamefont {Hansen}, \citenamefont
  {Petraru},\ and\ \citenamefont {Kohlstedt}}]{Ignatov2015}%
  \BibitemOpen
  \bibfield  {author} {\bibinfo {author} {\bibfnamefont {M.}~\bibnamefont
  {Ignatov}}, \bibinfo {author} {\bibfnamefont {M.}~\bibnamefont {Ziegler}},
  \bibinfo {author} {\bibfnamefont {M.}~\bibnamefont {Hansen}}, \bibinfo
  {author} {\bibfnamefont {A.}~\bibnamefont {Petraru}},\ and\ \bibinfo {author}
  {\bibfnamefont {H.}~\bibnamefont {Kohlstedt}},\ }\bibfield  {title} {\bibinfo
  {title} {A memristive spiking neuron with firing rate coding},\ }\href
  {https://doi.org/10.3389/fnins.2015.00376} {\bibfield  {journal} {\bibinfo
  {journal} {Frontiers in Neuroscience}\ }\textbf {\bibinfo {volume} {9}},\
  \bibinfo {pages} {376} (\bibinfo {year} {2015})}\BibitemShut {NoStop}%
\bibitem [{\citenamefont {Moon}\ \emph {et~al.}(2015)\citenamefont {Moon},
  \citenamefont {Cha}, \citenamefont {Park}, \citenamefont {Gi}, \citenamefont
  {Chu}, \citenamefont {Baek}, \citenamefont {Lee}, \citenamefont {Oh},\ and\
  \citenamefont {Hwang}}]{Moon2015}%
  \BibitemOpen
  \bibfield  {author} {\bibinfo {author} {\bibfnamefont {K.}~\bibnamefont
  {Moon}}, \bibinfo {author} {\bibfnamefont {E.}~\bibnamefont {Cha}}, \bibinfo
  {author} {\bibfnamefont {J.}~\bibnamefont {Park}}, \bibinfo {author}
  {\bibfnamefont {S.}~\bibnamefont {Gi}}, \bibinfo {author} {\bibfnamefont
  {M.}~\bibnamefont {Chu}}, \bibinfo {author} {\bibfnamefont {K.}~\bibnamefont
  {Baek}}, \bibinfo {author} {\bibfnamefont {B.}~\bibnamefont {Lee}}, \bibinfo
  {author} {\bibfnamefont {S.}~\bibnamefont {Oh}},\ and\ \bibinfo {author}
  {\bibfnamefont {H.}~\bibnamefont {Hwang}},\ }\bibfield  {title} {\bibinfo
  {title} {High density neuromorphic system with
  {M}o/{P}r$_{0.7}${C}a$_{0.3}${M}n{O}$_{3}$ synapse and {N}b{O}$_{2}$ {IMT}
  oscillator neuron},\ }in\ \href {https://doi.org/10.1109/IEDM.2015.7409721}
  {\emph {\bibinfo {booktitle} {2015 IEEE International Electron Devices
  Meeting (IEDM)}}}\ (\bibinfo  {publisher} {IEEE},\ \bibinfo {year} {2015})\
  pp.\ \bibinfo {pages} {17.6.1--17.6.4}\BibitemShut {NoStop}%
\bibitem [{\citenamefont {Zhou}\ and\ \citenamefont
  {Ramanathan}(2015)}]{YouZhou2015}%
  \BibitemOpen
  \bibfield  {author} {\bibinfo {author} {\bibfnamefont {Y.}~\bibnamefont
  {Zhou}}\ and\ \bibinfo {author} {\bibfnamefont {S.}~\bibnamefont
  {Ramanathan}},\ }\bibfield  {title} {\bibinfo {title} {Mott {M}emory and
  {N}euromorphic {D}evices},\ }\href
  {https://doi.org/10.1109/JPROC.2015.2431914} {\bibfield  {journal} {\bibinfo
  {journal} {Proceedings of the IEEE}\ }\textbf {\bibinfo {volume} {103}},\
  \bibinfo {pages} {1289} (\bibinfo {year} {2015})}\BibitemShut {NoStop}%
\bibitem [{\citenamefont {Prezioso}\ \emph {et~al.}(2015)\citenamefont
  {Prezioso}, \citenamefont {Merrikh-Bayat}, \citenamefont {Hoskins},
  \citenamefont {Adam}, \citenamefont {Likharev},\ and\ \citenamefont
  {Strukov}}]{Prezioso2015}%
  \BibitemOpen
  \bibfield  {author} {\bibinfo {author} {\bibfnamefont {M.}~\bibnamefont
  {Prezioso}}, \bibinfo {author} {\bibfnamefont {F.}~\bibnamefont
  {Merrikh-Bayat}}, \bibinfo {author} {\bibfnamefont {B.~D.}\ \bibnamefont
  {Hoskins}}, \bibinfo {author} {\bibfnamefont {G.~C.}\ \bibnamefont {Adam}},
  \bibinfo {author} {\bibfnamefont {K.~K.}\ \bibnamefont {Likharev}},\ and\
  \bibinfo {author} {\bibfnamefont {D.~B.}\ \bibnamefont {Strukov}},\
  }\bibfield  {title} {\bibinfo {title} {Training and operation of an
  integrated neuromorphic network based on metal-oxide memristors},\ }\href
  {https://doi.org/10.1038/nature14441} {\bibfield  {journal} {\bibinfo
  {journal} {Nature}\ }\textbf {\bibinfo {volume} {521}},\ \bibinfo {pages}
  {61} (\bibinfo {year} {2015})}\BibitemShut {NoStop}%
\bibitem [{\citenamefont {Kumar}\ \emph
  {et~al.}(2017{\natexlab{a}})\citenamefont {Kumar}, \citenamefont {Strachan},\
  and\ \citenamefont {Williams}}]{Kumar2017}%
  \BibitemOpen
  \bibfield  {author} {\bibinfo {author} {\bibfnamefont {S.}~\bibnamefont
  {Kumar}}, \bibinfo {author} {\bibfnamefont {J.~P.}\ \bibnamefont
  {Strachan}},\ and\ \bibinfo {author} {\bibfnamefont {R.~S.}\ \bibnamefont
  {Williams}},\ }\bibfield  {title} {\bibinfo {title} {Chaotic dynamics in
  nanoscale {N}b{O}$_{2}$ {M}ott memristors for analogue computing},\ }\href
  {https://doi.org/10.1038/nature23307} {\bibfield  {journal} {\bibinfo
  {journal} {Nature}\ }\textbf {\bibinfo {volume} {548}},\ \bibinfo {pages}
  {318} (\bibinfo {year} {2017}{\natexlab{a}})}\BibitemShut {NoStop}%
\bibitem [{\citenamefont {Stoliar}\ \emph {et~al.}(2017)\citenamefont
  {Stoliar}, \citenamefont {Tranchant}, \citenamefont {Corraze}, \citenamefont
  {Janod}, \citenamefont {Besland}, \citenamefont {Tesler}, \citenamefont
  {Rozenberg},\ and\ \citenamefont {Cario}}]{Stoliar2017}%
  \BibitemOpen
  \bibfield  {author} {\bibinfo {author} {\bibfnamefont {P.}~\bibnamefont
  {Stoliar}}, \bibinfo {author} {\bibfnamefont {J.}~\bibnamefont {Tranchant}},
  \bibinfo {author} {\bibfnamefont {B.}~\bibnamefont {Corraze}}, \bibinfo
  {author} {\bibfnamefont {E.}~\bibnamefont {Janod}}, \bibinfo {author}
  {\bibfnamefont {M.-P.}\ \bibnamefont {Besland}}, \bibinfo {author}
  {\bibfnamefont {F.}~\bibnamefont {Tesler}}, \bibinfo {author} {\bibfnamefont
  {M.}~\bibnamefont {Rozenberg}},\ and\ \bibinfo {author} {\bibfnamefont
  {L.}~\bibnamefont {Cario}},\ }\bibfield  {title} {\bibinfo {title} {A
  {L}eaky-{I}ntegrate-and-{F}ire {N}euron {A}nalog {R}ealized with a {M}ott
  {I}nsulator},\ }\href {https://doi.org/10.1002/adfm.201604740} {\bibfield
  {journal} {\bibinfo  {journal} {Advanced Functional Materials}\ }\textbf
  {\bibinfo {volume} {27}},\ \bibinfo {pages} {1604740} (\bibinfo {year}
  {2017})}\BibitemShut {NoStop}%
\bibitem [{\citenamefont {Yi}\ \emph {et~al.}(2018)\citenamefont {Yi},
  \citenamefont {Tsang}, \citenamefont {Lam}, \citenamefont {Bai},
  \citenamefont {Crowell},\ and\ \citenamefont {Flores}}]{Yi2018}%
  \BibitemOpen
  \bibfield  {author} {\bibinfo {author} {\bibfnamefont {W.}~\bibnamefont
  {Yi}}, \bibinfo {author} {\bibfnamefont {K.~K.}\ \bibnamefont {Tsang}},
  \bibinfo {author} {\bibfnamefont {S.~K.}\ \bibnamefont {Lam}}, \bibinfo
  {author} {\bibfnamefont {X.}~\bibnamefont {Bai}}, \bibinfo {author}
  {\bibfnamefont {J.~A.}\ \bibnamefont {Crowell}},\ and\ \bibinfo {author}
  {\bibfnamefont {E.~A.}\ \bibnamefont {Flores}},\ }\bibfield  {title}
  {\bibinfo {title} {Biological plausibility and stochasticity in scalable
  {VO}$_{2}$ active memristor neurons},\ }\href
  {https://doi.org/10.1038/s41467-018-07052-w} {\bibfield  {journal} {\bibinfo
  {journal} {Nature Communications}\ }\textbf {\bibinfo {volume} {9}},\
  \bibinfo {pages} {4661} (\bibinfo {year} {2018})}\BibitemShut {NoStop}%
\bibitem [{\citenamefont {del Valle}\ \emph {et~al.}(2018)\citenamefont {del
  Valle}, \citenamefont {Ram\'{i}rez}, \citenamefont {Rozenberg},\ and\
  \citenamefont {Schuller}}]{delValle2018}%
  \BibitemOpen
  \bibfield  {author} {\bibinfo {author} {\bibfnamefont {J.}~\bibnamefont {del
  Valle}}, \bibinfo {author} {\bibfnamefont {J.~G.}\ \bibnamefont
  {Ram\'{i}rez}}, \bibinfo {author} {\bibfnamefont {M.~J.}\ \bibnamefont
  {Rozenberg}},\ and\ \bibinfo {author} {\bibfnamefont {I.~K.}\ \bibnamefont
  {Schuller}},\ }\bibfield  {title} {\bibinfo {title} {Challenges in materials
  and devices for resistive-switching-based neuromorphic computing},\ }\href
  {https://doi.org/10.1063/1.5047800} {\bibfield  {journal} {\bibinfo
  {journal} {Journal of Applied Physics}\ }\textbf {\bibinfo {volume} {124}},\
  \bibinfo {pages} {211101} (\bibinfo {year} {2018})}\BibitemShut {NoStop}%
\bibitem [{\citenamefont {Lin}\ \emph {et~al.}(2018)\citenamefont {Lin},
  \citenamefont {Guha},\ and\ \citenamefont {Ramanathan}}]{Lin2018}%
  \BibitemOpen
  \bibfield  {author} {\bibinfo {author} {\bibfnamefont {J.}~\bibnamefont
  {Lin}}, \bibinfo {author} {\bibfnamefont {S.}~\bibnamefont {Guha}},\ and\
  \bibinfo {author} {\bibfnamefont {S.}~\bibnamefont {Ramanathan}},\ }\bibfield
   {title} {\bibinfo {title} {Vanadium {D}ioxide {C}ircuits {E}mulate
  {N}eurological {D}isorders},\ }\href
  {https://doi.org/10.3389/fnins.2018.00856} {\bibfield  {journal} {\bibinfo
  {journal} {Frontiers in Neuroscience}\ }\textbf {\bibinfo {volume} {12}},\
  \bibinfo {pages} {1} (\bibinfo {year} {2018})}\BibitemShut {NoStop}%
\bibitem [{\citenamefont {Bohaichuk}\ \emph {et~al.}(2019)\citenamefont
  {Bohaichuk}, \citenamefont {Kumar}, \citenamefont {Pitner}, \citenamefont
  {McClellan}, \citenamefont {Jeong}, \citenamefont {Samant}, \citenamefont
  {Wong}, \citenamefont {Parkin}, \citenamefont {Williams},\ and\ \citenamefont
  {Pop}}]{Bohaichuk2019}%
  \BibitemOpen
  \bibfield  {author} {\bibinfo {author} {\bibfnamefont {S.~M.}\ \bibnamefont
  {Bohaichuk}}, \bibinfo {author} {\bibfnamefont {S.}~\bibnamefont {Kumar}},
  \bibinfo {author} {\bibfnamefont {G.}~\bibnamefont {Pitner}}, \bibinfo
  {author} {\bibfnamefont {C.~J.}\ \bibnamefont {McClellan}}, \bibinfo {author}
  {\bibfnamefont {J.}~\bibnamefont {Jeong}}, \bibinfo {author} {\bibfnamefont
  {M.~G.}\ \bibnamefont {Samant}}, \bibinfo {author} {\bibfnamefont {H.~S.}\
  \bibnamefont {Wong}}, \bibinfo {author} {\bibfnamefont {S.~S.}\ \bibnamefont
  {Parkin}}, \bibinfo {author} {\bibfnamefont {R.~S.}\ \bibnamefont
  {Williams}},\ and\ \bibinfo {author} {\bibfnamefont {E.}~\bibnamefont
  {Pop}},\ }\bibfield  {title} {\bibinfo {title} {Fast {S}piking of a {M}ott
  {VO}$_{2}$-{C}arbon {N}anotube {C}omposite {D}evice},\ }\href
  {https://doi.org/10.1021/acs.nanolett.9b01554} {\bibfield  {journal}
  {\bibinfo  {journal} {Nano Letters}\ }\textbf {\bibinfo {volume} {19}},\
  \bibinfo {pages} {6751} (\bibinfo {year} {2019})}\BibitemShut {NoStop}%
\bibitem [{\citenamefont {Salev}\ \emph {et~al.}(2019)\citenamefont {Salev},
  \citenamefont {Valle}, \citenamefont {Kalcheim},\ and\ \citenamefont
  {Schuller}}]{Salev2019}%
  \BibitemOpen
  \bibfield  {author} {\bibinfo {author} {\bibfnamefont {P.}~\bibnamefont
  {Salev}}, \bibinfo {author} {\bibfnamefont {J.~D.}\ \bibnamefont {Valle}},
  \bibinfo {author} {\bibfnamefont {Y.}~\bibnamefont {Kalcheim}},\ and\
  \bibinfo {author} {\bibfnamefont {I.~K.}\ \bibnamefont {Schuller}},\
  }\bibfield  {title} {\bibinfo {title} {Giant nonvolatile resistive switching
  in a mott oxide and ferroelectric hybrid},\ }\href
  {https://doi.org/10.1073/pnas.1822138116} {\bibfield  {journal} {\bibinfo
  {journal} {Proceedings of the National Academy of Sciences of the United
  States of America}\ }\textbf {\bibinfo {volume} {116}},\ \bibinfo {pages}
  {8798} (\bibinfo {year} {2019})}\BibitemShut {NoStop}%
\bibitem [{\citenamefont {Feldmann}\ \emph {et~al.}(2019)\citenamefont
  {Feldmann}, \citenamefont {Youngblood}, \citenamefont {Wright}, \citenamefont
  {Bhaskaran},\ and\ \citenamefont {Pernice}}]{Feldmann2019}%
  \BibitemOpen
  \bibfield  {author} {\bibinfo {author} {\bibfnamefont {J.}~\bibnamefont
  {Feldmann}}, \bibinfo {author} {\bibfnamefont {N.}~\bibnamefont
  {Youngblood}}, \bibinfo {author} {\bibfnamefont {C.~D.}\ \bibnamefont
  {Wright}}, \bibinfo {author} {\bibfnamefont {H.}~\bibnamefont {Bhaskaran}},\
  and\ \bibinfo {author} {\bibfnamefont {W.~H.}\ \bibnamefont {Pernice}},\
  }\bibfield  {title} {\bibinfo {title} {All-optical spiking neurosynaptic
  networks with self-learning capabilities},\ }\href
  {https://doi.org/10.1038/s41586-019-1157-8} {\bibfield  {journal} {\bibinfo
  {journal} {Nature}\ }\textbf {\bibinfo {volume} {569}},\ \bibinfo {pages}
  {208} (\bibinfo {year} {2019})}\BibitemShut {NoStop}%
\bibitem [{\citenamefont {del Valle}\ \emph {et~al.}(2020)\citenamefont {del
  Valle}, \citenamefont {Salev}, \citenamefont {Kalcheim},\ and\ \citenamefont
  {Schuller}}]{delValle2020}%
  \BibitemOpen
  \bibfield  {author} {\bibinfo {author} {\bibfnamefont {J.}~\bibnamefont {del
  Valle}}, \bibinfo {author} {\bibfnamefont {P.}~\bibnamefont {Salev}},
  \bibinfo {author} {\bibfnamefont {Y.}~\bibnamefont {Kalcheim}},\ and\
  \bibinfo {author} {\bibfnamefont {I.~K.}\ \bibnamefont {Schuller}},\
  }\bibfield  {title} {\bibinfo {title} {A caloritronics-based {M}ott
  neuristor},\ }\href {https://doi.org/10.1038/s41598-020-61176-y} {\bibfield
  {journal} {\bibinfo  {journal} {Scientific Reports}\ }\textbf {\bibinfo
  {volume} {10}},\ \bibinfo {pages} {4292} (\bibinfo {year}
  {2020})}\BibitemShut {NoStop}%
\bibitem [{\citenamefont {Kumar}\ \emph {et~al.}(2020)\citenamefont {Kumar},
  \citenamefont {Williams},\ and\ \citenamefont {Wang}}]{Kumar2020}%
  \BibitemOpen
  \bibfield  {author} {\bibinfo {author} {\bibfnamefont {S.}~\bibnamefont
  {Kumar}}, \bibinfo {author} {\bibfnamefont {R.~S.}\ \bibnamefont
  {Williams}},\ and\ \bibinfo {author} {\bibfnamefont {Z.}~\bibnamefont
  {Wang}},\ }\bibfield  {title} {\bibinfo {title} {Third-order nanocircuit
  elements for neuromorphic engineering},\ }\href
  {https://doi.org/10.1038/s41586-020-2735-5} {\bibfield  {journal} {\bibinfo
  {journal} {Nature}\ }\textbf {\bibinfo {volume} {585}},\ \bibinfo {pages}
  {518} (\bibinfo {year} {2020})}\BibitemShut {NoStop}%
\bibitem [{\citenamefont {Oh}\ \emph {et~al.}(2021)\citenamefont {Oh},
  \citenamefont {Shi}, \citenamefont {del Valle}, \citenamefont {Salev},
  \citenamefont {Lu}, \citenamefont {Huang}, \citenamefont {Kalcheim},
  \citenamefont {Schuller},\ and\ \citenamefont {Kuzum}}]{Oh2021}%
  \BibitemOpen
  \bibfield  {author} {\bibinfo {author} {\bibfnamefont {S.}~\bibnamefont
  {Oh}}, \bibinfo {author} {\bibfnamefont {Y.}~\bibnamefont {Shi}}, \bibinfo
  {author} {\bibfnamefont {J.}~\bibnamefont {del Valle}}, \bibinfo {author}
  {\bibfnamefont {P.}~\bibnamefont {Salev}}, \bibinfo {author} {\bibfnamefont
  {Y.}~\bibnamefont {Lu}}, \bibinfo {author} {\bibfnamefont {Z.}~\bibnamefont
  {Huang}}, \bibinfo {author} {\bibfnamefont {Y.}~\bibnamefont {Kalcheim}},
  \bibinfo {author} {\bibfnamefont {I.~K.}\ \bibnamefont {Schuller}},\ and\
  \bibinfo {author} {\bibfnamefont {D.}~\bibnamefont {Kuzum}},\ }\bibfield
  {title} {\bibinfo {title} {Energy-efficient {M}ott activation neuron for
  full-hardware implementation of neural networks},\ }\href
  {https://doi.org/10.1038/s41565-021-00874-8} {\bibfield  {journal} {\bibinfo
  {journal} {Nature Nanotechnology}\ }\textbf {\bibinfo {volume} {16}},\
  \bibinfo {pages} {680} (\bibinfo {year} {2021})}\BibitemShut {NoStop}%
\bibitem [{\citenamefont {Lin}\ and\ \citenamefont {Shen}(2023)}]{Lin2023}%
  \BibitemOpen
  \bibfield  {author} {\bibinfo {author} {\bibfnamefont {H.}~\bibnamefont
  {Lin}}\ and\ \bibinfo {author} {\bibfnamefont {Y.}~\bibnamefont {Shen}},\
  }\bibfield  {title} {\bibinfo {title} {A {VO}$_{2}$ {N}euristor {B}ased on
  {M}icrostrip {L}ine {C}oupling},\ }\href {https://doi.org/10.3390/mi14020337}
  {\bibfield  {journal} {\bibinfo  {journal} {Micromachines}\ }\textbf
  {\bibinfo {volume} {14}},\ \bibinfo {pages} {337} (\bibinfo {year}
  {2023})}\BibitemShut {NoStop}%
\bibitem [{\citenamefont {Schofield}\ \emph {et~al.}(2023)\citenamefont
  {Schofield}, \citenamefont {Bradicich}, \citenamefont {Gurrola},
  \citenamefont {Zhang}, \citenamefont {Brown}, \citenamefont {Pharr},
  \citenamefont {Shamberger},\ and\ \citenamefont {Banerjee}}]{Schofield2023}%
  \BibitemOpen
  \bibfield  {author} {\bibinfo {author} {\bibfnamefont {P.}~\bibnamefont
  {Schofield}}, \bibinfo {author} {\bibfnamefont {A.}~\bibnamefont
  {Bradicich}}, \bibinfo {author} {\bibfnamefont {R.~M.}\ \bibnamefont
  {Gurrola}}, \bibinfo {author} {\bibfnamefont {Y.}~\bibnamefont {Zhang}},
  \bibinfo {author} {\bibfnamefont {T.~D.}\ \bibnamefont {Brown}}, \bibinfo
  {author} {\bibfnamefont {M.}~\bibnamefont {Pharr}}, \bibinfo {author}
  {\bibfnamefont {P.~J.}\ \bibnamefont {Shamberger}},\ and\ \bibinfo {author}
  {\bibfnamefont {S.}~\bibnamefont {Banerjee}},\ }\bibfield  {title} {\bibinfo
  {title} {Harnessing the {M}etal-{I}nsulator {T}ransition of {VO}$_{2}$ in
  {N}euromorphic {C}omputing},\ }\href {https://doi.org/10.1002/adma.202205294}
  {\bibfield  {journal} {\bibinfo  {journal} {Advanced Materials}\ }\textbf
  {\bibinfo {volume} {35}},\ \bibinfo {pages} {2205294} (\bibinfo {year}
  {2023})}\BibitemShut {NoStop}%
\bibitem [{\citenamefont {Zimmers}\ \emph {et~al.}(2013)\citenamefont
  {Zimmers}, \citenamefont {Aigouy}, \citenamefont {Mortier}, \citenamefont
  {Sharoni}, \citenamefont {Wang}, \citenamefont {West}, \citenamefont
  {Ramirez},\ and\ \citenamefont {Schuller}}]{Zimmers2013}%
  \BibitemOpen
  \bibfield  {author} {\bibinfo {author} {\bibfnamefont {A.}~\bibnamefont
  {Zimmers}}, \bibinfo {author} {\bibfnamefont {L.}~\bibnamefont {Aigouy}},
  \bibinfo {author} {\bibfnamefont {M.}~\bibnamefont {Mortier}}, \bibinfo
  {author} {\bibfnamefont {A.}~\bibnamefont {Sharoni}}, \bibinfo {author}
  {\bibfnamefont {S.}~\bibnamefont {Wang}}, \bibinfo {author} {\bibfnamefont
  {K.~G.}\ \bibnamefont {West}}, \bibinfo {author} {\bibfnamefont {J.~G.}\
  \bibnamefont {Ramirez}},\ and\ \bibinfo {author} {\bibfnamefont {I.~K.}\
  \bibnamefont {Schuller}},\ }\bibfield  {title} {\bibinfo {title} {Role of
  {T}hermal {H}eating on the {V}oltage {I}nduced {I}nsulator-{M}etal
  {T}ransition in {VO}$_{2}$},\ }\href
  {https://doi.org/10.1103/PhysRevLett.110.056601} {\bibfield  {journal}
  {\bibinfo  {journal} {Physical Review Letters}\ }\textbf {\bibinfo {volume}
  {110}},\ \bibinfo {pages} {056601} (\bibinfo {year} {2013})}\BibitemShut
  {NoStop}%
\bibitem [{\citenamefont {Shukla}\ \emph {et~al.}(2014)\citenamefont {Shukla},
  \citenamefont {Joshi}, \citenamefont {Dasgupta}, \citenamefont {Borisov},
  \citenamefont {Lederman},\ and\ \citenamefont {Datta}}]{Shukla2014}%
  \BibitemOpen
  \bibfield  {author} {\bibinfo {author} {\bibfnamefont {N.}~\bibnamefont
  {Shukla}}, \bibinfo {author} {\bibfnamefont {T.}~\bibnamefont {Joshi}},
  \bibinfo {author} {\bibfnamefont {S.}~\bibnamefont {Dasgupta}}, \bibinfo
  {author} {\bibfnamefont {P.}~\bibnamefont {Borisov}}, \bibinfo {author}
  {\bibfnamefont {D.}~\bibnamefont {Lederman}},\ and\ \bibinfo {author}
  {\bibfnamefont {S.}~\bibnamefont {Datta}},\ }\bibfield  {title} {\bibinfo
  {title} {Electrically induced insulator to metal transition in epitaxial
  {S}m{N}i{O}$_{3}$ thin films},\ }\href {https://doi.org/10.1063/1.4890329}
  {\bibfield  {journal} {\bibinfo  {journal} {Applied Physics Letters}\
  }\textbf {\bibinfo {volume} {105}},\ \bibinfo {pages} {012108} (\bibinfo
  {year} {2014})}\BibitemShut {NoStop}%
\bibitem [{\citenamefont {Liu}\ \emph {et~al.}(2016)\citenamefont {Liu},
  \citenamefont {Li}, \citenamefont {Nandi}, \citenamefont {Venkatachalam},\
  and\ \citenamefont {Elliman}}]{Liu2016}%
  \BibitemOpen
  \bibfield  {author} {\bibinfo {author} {\bibfnamefont {X.}~\bibnamefont
  {Liu}}, \bibinfo {author} {\bibfnamefont {S.}~\bibnamefont {Li}}, \bibinfo
  {author} {\bibfnamefont {S.~K.}\ \bibnamefont {Nandi}}, \bibinfo {author}
  {\bibfnamefont {D.~K.}\ \bibnamefont {Venkatachalam}},\ and\ \bibinfo
  {author} {\bibfnamefont {R.~G.}\ \bibnamefont {Elliman}},\ }\bibfield
  {title} {\bibinfo {title} {Threshold switching and electrical
  self-oscillation in niobium oxide films},\ }\href
  {https://doi.org/10.1063/1.4963288} {\bibfield  {journal} {\bibinfo
  {journal} {Journal of Applied Physics}\ }\textbf {\bibinfo {volume} {120}},\
  \bibinfo {pages} {10} (\bibinfo {year} {2016})}\BibitemShut {NoStop}%
\bibitem [{\citenamefont {del Valle}\ \emph
  {et~al.}(2021{\natexlab{a}})\citenamefont {del Valle}, \citenamefont {Rocco},
  \citenamefont {Dom\'{i}nguez}, \citenamefont {Fowlie}, \citenamefont
  {Gariglio}, \citenamefont {Rozenberg},\ and\ \citenamefont
  {Triscone}}]{delValle2021}%
  \BibitemOpen
  \bibfield  {author} {\bibinfo {author} {\bibfnamefont {J.}~\bibnamefont {del
  Valle}}, \bibinfo {author} {\bibfnamefont {R.}~\bibnamefont {Rocco}},
  \bibinfo {author} {\bibfnamefont {C.}~\bibnamefont {Dom\'{i}nguez}}, \bibinfo
  {author} {\bibfnamefont {J.}~\bibnamefont {Fowlie}}, \bibinfo {author}
  {\bibfnamefont {S.}~\bibnamefont {Gariglio}}, \bibinfo {author}
  {\bibfnamefont {M.~J.}\ \bibnamefont {Rozenberg}},\ and\ \bibinfo {author}
  {\bibfnamefont {J.-M.}\ \bibnamefont {Triscone}},\ }\bibfield  {title}
  {\bibinfo {title} {Dynamics of the electrically induced insulator-to-metal
  transition in rare-earth nickelates},\ }\href
  {https://doi.org/10.1103/PhysRevB.104.165141} {\bibfield  {journal} {\bibinfo
   {journal} {Physical Review B}\ }\textbf {\bibinfo {volume} {104}},\ \bibinfo
  {pages} {165141} (\bibinfo {year} {2021}{\natexlab{a}})}\BibitemShut
  {NoStop}%
\bibitem [{\citenamefont {Salev}\ \emph {et~al.}(2021)\citenamefont {Salev},
  \citenamefont {Fratino}, \citenamefont {Sasaki}, \citenamefont {Berkoun},
  \citenamefont {del Valle}, \citenamefont {Kalcheim}, \citenamefont
  {Takamura}, \citenamefont {Rozenberg},\ and\ \citenamefont
  {Schuller}}]{Salev2021}%
  \BibitemOpen
  \bibfield  {author} {\bibinfo {author} {\bibfnamefont {P.}~\bibnamefont
  {Salev}}, \bibinfo {author} {\bibfnamefont {L.}~\bibnamefont {Fratino}},
  \bibinfo {author} {\bibfnamefont {D.}~\bibnamefont {Sasaki}}, \bibinfo
  {author} {\bibfnamefont {R.}~\bibnamefont {Berkoun}}, \bibinfo {author}
  {\bibfnamefont {J.}~\bibnamefont {del Valle}}, \bibinfo {author}
  {\bibfnamefont {Y.}~\bibnamefont {Kalcheim}}, \bibinfo {author}
  {\bibfnamefont {Y.}~\bibnamefont {Takamura}}, \bibinfo {author}
  {\bibfnamefont {M.}~\bibnamefont {Rozenberg}},\ and\ \bibinfo {author}
  {\bibfnamefont {I.~K.}\ \bibnamefont {Schuller}},\ }\bibfield  {title}
  {\bibinfo {title} {Transverse barrier formation by electrical triggering of a
  metal-to-insulator transition},\ }\href
  {https://doi.org/10.1038/s41467-021-25802-1} {\bibfield  {journal} {\bibinfo
  {journal} {Nature Communications}\ }\textbf {\bibinfo {volume} {12}},\
  \bibinfo {pages} {5499} (\bibinfo {year} {2021})}\BibitemShut {NoStop}%
\bibitem [{\citenamefont {Rocco}\ \emph {et~al.}(2022)\citenamefont {Rocco},
  \citenamefont {del Valle}, \citenamefont {Navarro}, \citenamefont {Salev},
  \citenamefont {Schuller},\ and\ \citenamefont {Rozenberg}}]{Rocco2022}%
  \BibitemOpen
  \bibfield  {author} {\bibinfo {author} {\bibfnamefont {R.}~\bibnamefont
  {Rocco}}, \bibinfo {author} {\bibfnamefont {J.}~\bibnamefont {del Valle}},
  \bibinfo {author} {\bibfnamefont {H.}~\bibnamefont {Navarro}}, \bibinfo
  {author} {\bibfnamefont {P.}~\bibnamefont {Salev}}, \bibinfo {author}
  {\bibfnamefont {I.~K.}\ \bibnamefont {Schuller}},\ and\ \bibinfo {author}
  {\bibfnamefont {M.}~\bibnamefont {Rozenberg}},\ }\bibfield  {title} {\bibinfo
  {title} {Exponential {E}scape {R}ate of {F}ilamentary {I}ncubation in {M}ott
  {S}piking {N}eurons},\ }\href
  {https://doi.org/10.1103/PhysRevApplied.17.024028} {\bibfield  {journal}
  {\bibinfo  {journal} {Physical Review Applied}\ }\textbf {\bibinfo {volume}
  {17}},\ \bibinfo {pages} {024028} (\bibinfo {year} {2022})}\BibitemShut
  {NoStop}%
\bibitem [{\citenamefont {Adda}\ \emph {et~al.}(2022)\citenamefont {Adda},
  \citenamefont {Lee}, \citenamefont {Kalcheim}, \citenamefont {Salev},
  \citenamefont {Rocco}, \citenamefont {Vargas}, \citenamefont {Ghazikhanian},
  \citenamefont {Li}, \citenamefont {Albright}, \citenamefont {Rozenberg},\
  and\ \citenamefont {Schuller}}]{Adda2022}%
  \BibitemOpen
  \bibfield  {author} {\bibinfo {author} {\bibfnamefont {C.}~\bibnamefont
  {Adda}}, \bibinfo {author} {\bibfnamefont {M.-H.}\ \bibnamefont {Lee}},
  \bibinfo {author} {\bibfnamefont {Y.}~\bibnamefont {Kalcheim}}, \bibinfo
  {author} {\bibfnamefont {P.}~\bibnamefont {Salev}}, \bibinfo {author}
  {\bibfnamefont {R.}~\bibnamefont {Rocco}}, \bibinfo {author} {\bibfnamefont
  {N.~M.}\ \bibnamefont {Vargas}}, \bibinfo {author} {\bibfnamefont
  {N.}~\bibnamefont {Ghazikhanian}}, \bibinfo {author} {\bibfnamefont {C.-P.}\
  \bibnamefont {Li}}, \bibinfo {author} {\bibfnamefont {G.}~\bibnamefont
  {Albright}}, \bibinfo {author} {\bibfnamefont {M.}~\bibnamefont
  {Rozenberg}},\ and\ \bibinfo {author} {\bibfnamefont {I.~K.}\ \bibnamefont
  {Schuller}},\ }\bibfield  {title} {\bibinfo {title} {Direct {O}bservation of
  the {E}lectrically {T}riggered {I}nsulator-{M}etal {T}ransition in
  {V}$_{3}${O}$_{5}$ {F}ar below the {T}ransition {T}emperature},\ }\href
  {https://doi.org/10.1103/PhysRevX.12.011025} {\bibfield  {journal} {\bibinfo
  {journal} {Physical Review X}\ }\textbf {\bibinfo {volume} {12}},\ \bibinfo
  {pages} {011025} (\bibinfo {year} {2022})}\BibitemShut {NoStop}%
\bibitem [{\citenamefont {Luibrand}\ \emph {et~al.}(2023)\citenamefont
  {Luibrand}, \citenamefont {Bercher}, \citenamefont {Rocco}, \citenamefont
  {Tahouni-Bonab}, \citenamefont {Varbaro}, \citenamefont {Rischau},
  \citenamefont {Dom\'{i}nguez}, \citenamefont {Zhou}, \citenamefont {Luo},
  \citenamefont {Bag}, \citenamefont {Fratino}, \citenamefont {Kleiner},
  \citenamefont {Gariglio}, \citenamefont {Koelle}, \citenamefont {Triscone},
  \citenamefont {Rozenberg}, \citenamefont {Kuzmenko}, \citenamefont
  {Gu\'{e}non},\ and\ \citenamefont {del Valle}}]{Luibrand2023}%
  \BibitemOpen
  \bibfield  {author} {\bibinfo {author} {\bibfnamefont {T.}~\bibnamefont
  {Luibrand}}, \bibinfo {author} {\bibfnamefont {A.}~\bibnamefont {Bercher}},
  \bibinfo {author} {\bibfnamefont {R.}~\bibnamefont {Rocco}}, \bibinfo
  {author} {\bibfnamefont {F.}~\bibnamefont {Tahouni-Bonab}}, \bibinfo {author}
  {\bibfnamefont {L.}~\bibnamefont {Varbaro}}, \bibinfo {author} {\bibfnamefont
  {C.~W.}\ \bibnamefont {Rischau}}, \bibinfo {author} {\bibfnamefont
  {C.}~\bibnamefont {Dom\'{i}nguez}}, \bibinfo {author} {\bibfnamefont
  {Y.}~\bibnamefont {Zhou}}, \bibinfo {author} {\bibfnamefont {W.}~\bibnamefont
  {Luo}}, \bibinfo {author} {\bibfnamefont {S.}~\bibnamefont {Bag}}, \bibinfo
  {author} {\bibfnamefont {L.}~\bibnamefont {Fratino}}, \bibinfo {author}
  {\bibfnamefont {R.}~\bibnamefont {Kleiner}}, \bibinfo {author} {\bibfnamefont
  {S.}~\bibnamefont {Gariglio}}, \bibinfo {author} {\bibfnamefont
  {D.}~\bibnamefont {Koelle}}, \bibinfo {author} {\bibfnamefont {J.-M.}\
  \bibnamefont {Triscone}}, \bibinfo {author} {\bibfnamefont {M.~J.}\
  \bibnamefont {Rozenberg}}, \bibinfo {author} {\bibfnamefont {A.~B.}\
  \bibnamefont {Kuzmenko}}, \bibinfo {author} {\bibfnamefont {S.}~\bibnamefont
  {Gu\'{e}non}},\ and\ \bibinfo {author} {\bibfnamefont {J.}~\bibnamefont {del
  Valle}},\ }\bibfield  {title} {\bibinfo {title} {Characteristic length scales
  of the electrically induced insulator-to-metal transition},\ }\href
  {https://doi.org/10.1103/PhysRevResearch.5.013108} {\bibfield  {journal}
  {\bibinfo  {journal} {Physical Review Research}\ }\textbf {\bibinfo {volume}
  {5}},\ \bibinfo {pages} {013108} (\bibinfo {year} {2023})}\BibitemShut
  {NoStop}%
\bibitem [{\citenamefont {Rischau}\ \emph {et~al.}(2024)\citenamefont
  {Rischau}, \citenamefont {Gariglio}, \citenamefont {Triscone},\ and\
  \citenamefont {del Valle}}]{Rischau2024}%
  \BibitemOpen
  \bibfield  {author} {\bibinfo {author} {\bibfnamefont {C.~W.}\ \bibnamefont
  {Rischau}}, \bibinfo {author} {\bibfnamefont {S.}~\bibnamefont {Gariglio}},
  \bibinfo {author} {\bibfnamefont {J.-M.}\ \bibnamefont {Triscone}},\ and\
  \bibinfo {author} {\bibfnamefont {J.}~\bibnamefont {del Valle}},\ }\bibfield
  {title} {\bibinfo {title} {Resistive switching of {VO}$_{2}$ films grown on a
  thermal insulator},\ }\href
  {https://doi.org/10.1103/PhysRevApplied.22.014021} {\bibfield  {journal}
  {\bibinfo  {journal} {Physical Review Applied}\ }\textbf {\bibinfo {volume}
  {22}},\ \bibinfo {pages} {014021} (\bibinfo {year} {2024})}\BibitemShut
  {NoStop}%
\bibitem [{\citenamefont {Valmianski}\ \emph {et~al.}(2018)\citenamefont
  {Valmianski}, \citenamefont {Wang}, \citenamefont {Wang}, \citenamefont
  {Ramirez}, \citenamefont {Gu\'{e}non},\ and\ \citenamefont
  {Schuller}}]{Valmianski2018}%
  \BibitemOpen
  \bibfield  {author} {\bibinfo {author} {\bibfnamefont {I.}~\bibnamefont
  {Valmianski}}, \bibinfo {author} {\bibfnamefont {P.~Y.}\ \bibnamefont
  {Wang}}, \bibinfo {author} {\bibfnamefont {S.}~\bibnamefont {Wang}}, \bibinfo
  {author} {\bibfnamefont {J.~G.}\ \bibnamefont {Ramirez}}, \bibinfo {author}
  {\bibfnamefont {S.}~\bibnamefont {Gu\'{e}non}},\ and\ \bibinfo {author}
  {\bibfnamefont {I.~K.}\ \bibnamefont {Schuller}},\ }\bibfield  {title}
  {\bibinfo {title} {Origin of the current-driven breakdown in vanadium oxides:
  {T}hermal versus electronic},\ }\href
  {https://doi.org/10.1103/PhysRevB.98.195144} {\bibfield  {journal} {\bibinfo
  {journal} {Physical Review B}\ }\textbf {\bibinfo {volume} {98}},\ \bibinfo
  {pages} {195144} (\bibinfo {year} {2018})}\BibitemShut {NoStop}%
\bibitem [{\citenamefont {Kalcheim}\ \emph {et~al.}(2020)\citenamefont
  {Kalcheim}, \citenamefont {Camjayi}, \citenamefont {del Valle}, \citenamefont
  {Salev}, \citenamefont {Rozenberg},\ and\ \citenamefont
  {Schuller}}]{Kalcheim2020}%
  \BibitemOpen
  \bibfield  {author} {\bibinfo {author} {\bibfnamefont {Y.}~\bibnamefont
  {Kalcheim}}, \bibinfo {author} {\bibfnamefont {A.}~\bibnamefont {Camjayi}},
  \bibinfo {author} {\bibfnamefont {J.}~\bibnamefont {del Valle}}, \bibinfo
  {author} {\bibfnamefont {P.}~\bibnamefont {Salev}}, \bibinfo {author}
  {\bibfnamefont {M.}~\bibnamefont {Rozenberg}},\ and\ \bibinfo {author}
  {\bibfnamefont {I.~K.}\ \bibnamefont {Schuller}},\ }\bibfield  {title}
  {\bibinfo {title} {Non-thermal resistive switching in {M}ott insulator
  nanowires},\ }\href {https://doi.org/10.1038/s41467-020-16752-1} {\bibfield
  {journal} {\bibinfo  {journal} {Nature Communications}\ }\textbf {\bibinfo
  {volume} {11}},\ \bibinfo {pages} {2985} (\bibinfo {year}
  {2020})}\BibitemShut {NoStop}%
\bibitem [{\citenamefont {Sahoo}\ \emph {et~al.}(2023)\citenamefont {Sahoo},
  \citenamefont {Jana}, \citenamefont {Yadav}, \citenamefont {Rawat},
  \citenamefont {Phase},\ and\ \citenamefont {Choudhary}}]{Sahoo2023}%
  \BibitemOpen
  \bibfield  {author} {\bibinfo {author} {\bibfnamefont {S.}~\bibnamefont
  {Sahoo}}, \bibinfo {author} {\bibfnamefont {A.}~\bibnamefont {Jana}},
  \bibinfo {author} {\bibfnamefont {S.}~\bibnamefont {Yadav}}, \bibinfo
  {author} {\bibfnamefont {R.}~\bibnamefont {Rawat}}, \bibinfo {author}
  {\bibfnamefont {D.}~\bibnamefont {Phase}},\ and\ \bibinfo {author}
  {\bibfnamefont {R.}~\bibnamefont {Choudhary}},\ }\bibfield  {title} {\bibinfo
  {title} {Electrically {I}nduced {N}onthermal {M}emristive {S}witching in
  {V}$_{2}${O}$_{3}$/{S}i {T}hin {F}ilm},\ }\href
  {https://doi.org/10.1103/PhysRevApplied.20.024055} {\bibfield  {journal}
  {\bibinfo  {journal} {Physical Review Applied}\ }\textbf {\bibinfo {volume}
  {20}},\ \bibinfo {pages} {024055} (\bibinfo {year} {2023})}\BibitemShut
  {NoStop}%
\bibitem [{\citenamefont {del Valle}\ \emph
  {et~al.}(2021{\natexlab{b}})\citenamefont {del Valle}, \citenamefont
  {Vargas}, \citenamefont {Rocco}, \citenamefont {Salev}, \citenamefont
  {Kalcheim}, \citenamefont {Lapa}, \citenamefont {Adda}, \citenamefont {Lee},
  \citenamefont {Wang}, \citenamefont {Fratino}, \citenamefont {Rozenberg},\
  and\ \citenamefont {Schuller}}]{delValle2021_2}%
  \BibitemOpen
  \bibfield  {author} {\bibinfo {author} {\bibfnamefont {J.}~\bibnamefont {del
  Valle}}, \bibinfo {author} {\bibfnamefont {N.~M.}\ \bibnamefont {Vargas}},
  \bibinfo {author} {\bibfnamefont {R.}~\bibnamefont {Rocco}}, \bibinfo
  {author} {\bibfnamefont {P.}~\bibnamefont {Salev}}, \bibinfo {author}
  {\bibfnamefont {Y.}~\bibnamefont {Kalcheim}}, \bibinfo {author}
  {\bibfnamefont {P.~N.}\ \bibnamefont {Lapa}}, \bibinfo {author}
  {\bibfnamefont {C.}~\bibnamefont {Adda}}, \bibinfo {author} {\bibfnamefont
  {M.-H.}\ \bibnamefont {Lee}}, \bibinfo {author} {\bibfnamefont {P.~Y.}\
  \bibnamefont {Wang}}, \bibinfo {author} {\bibfnamefont {L.}~\bibnamefont
  {Fratino}}, \bibinfo {author} {\bibfnamefont {M.~J.}\ \bibnamefont
  {Rozenberg}},\ and\ \bibinfo {author} {\bibfnamefont {I.~K.}\ \bibnamefont
  {Schuller}},\ }\bibfield  {title} {\bibinfo {title} {Spatiotemporal
  characterization of the field-induced insulator-to-metal transition},\ }\href
  {https://doi.org/10.1126/science.abd9088} {\bibfield  {journal} {\bibinfo
  {journal} {Science}\ }\textbf {\bibinfo {volume} {373}},\ \bibinfo {pages}
  {907} (\bibinfo {year} {2021}{\natexlab{b}})}\BibitemShut {NoStop}%
\bibitem [{\citenamefont {Brockman}\ \emph {et~al.}(2014)\citenamefont
  {Brockman}, \citenamefont {Gao}, \citenamefont {Hughes}, \citenamefont
  {Rettner}, \citenamefont {Samant}, \citenamefont {Roche},\ and\ \citenamefont
  {Parkin}}]{Brockman2014}%
  \BibitemOpen
  \bibfield  {author} {\bibinfo {author} {\bibfnamefont {J.~S.}\ \bibnamefont
  {Brockman}}, \bibinfo {author} {\bibfnamefont {L.}~\bibnamefont {Gao}},
  \bibinfo {author} {\bibfnamefont {B.}~\bibnamefont {Hughes}}, \bibinfo
  {author} {\bibfnamefont {C.~T.}\ \bibnamefont {Rettner}}, \bibinfo {author}
  {\bibfnamefont {M.~G.}\ \bibnamefont {Samant}}, \bibinfo {author}
  {\bibfnamefont {K.~P.}\ \bibnamefont {Roche}},\ and\ \bibinfo {author}
  {\bibfnamefont {S.~S.}\ \bibnamefont {Parkin}},\ }\bibfield  {title}
  {\bibinfo {title} {Subnanosecond incubation times for electric-field-induced
  metallization of a correlated electron oxide},\ }\href
  {https://doi.org/10.1038/nnano.2014.71} {\bibfield  {journal} {\bibinfo
  {journal} {Nature Nanotechnology}\ }\textbf {\bibinfo {volume} {9}},\
  \bibinfo {pages} {453} (\bibinfo {year} {2014})}\BibitemShut {NoStop}%
\bibitem [{\citenamefont {Huang}\ \emph {et~al.}(2014)\citenamefont {Huang},
  \citenamefont {Luo}, \citenamefont {Yang}, \citenamefont {Yun}, \citenamefont
  {Yang}, \citenamefont {Meng}, \citenamefont {Wang}, \citenamefont {Hu},
  \citenamefont {Bao}, \citenamefont {Lu},\ and\ \citenamefont
  {Gao}}]{Huang2014}%
  \BibitemOpen
  \bibfield  {author} {\bibinfo {author} {\bibfnamefont {H.}~\bibnamefont
  {Huang}}, \bibinfo {author} {\bibfnamefont {Z.}~\bibnamefont {Luo}}, \bibinfo
  {author} {\bibfnamefont {Y.}~\bibnamefont {Yang}}, \bibinfo {author}
  {\bibfnamefont {Y.}~\bibnamefont {Yun}}, \bibinfo {author} {\bibfnamefont
  {M.}~\bibnamefont {Yang}}, \bibinfo {author} {\bibfnamefont {D.}~\bibnamefont
  {Meng}}, \bibinfo {author} {\bibfnamefont {H.}~\bibnamefont {Wang}}, \bibinfo
  {author} {\bibfnamefont {S.}~\bibnamefont {Hu}}, \bibinfo {author}
  {\bibfnamefont {J.}~\bibnamefont {Bao}}, \bibinfo {author} {\bibfnamefont
  {Y.}~\bibnamefont {Lu}},\ and\ \bibinfo {author} {\bibfnamefont
  {C.}~\bibnamefont {Gao}},\ }\bibfield  {title} {\bibinfo {title} {{DC}
  current induced metal-insulator transition in epitaxial
  {S}m$_{0.6}${N}d$_{0.4}${N}i{O}$_{3}$/{L}a{A}l{O}$_{3}$ thin film},\ }\href
  {https://doi.org/10.1063/1.4874642} {\bibfield  {journal} {\bibinfo
  {journal} {AIP Advances}\ }\textbf {\bibinfo {volume} {4}},\ \bibinfo {pages}
  {057102} (\bibinfo {year} {2014})}\BibitemShut {NoStop}%
\bibitem [{\citenamefont {del Valle}\ \emph {et~al.}(2019)\citenamefont {del
  Valle}, \citenamefont {Salev}, \citenamefont {Tesler}, \citenamefont
  {Vargas}, \citenamefont {Kalcheim}, \citenamefont {Wang}, \citenamefont
  {Trastoy}, \citenamefont {Lee}, \citenamefont {Kassabian}, \citenamefont
  {Ram\'{i}rez}, \citenamefont {Rozenberg},\ and\ \citenamefont
  {Schuller}}]{delValle2019}%
  \BibitemOpen
  \bibfield  {author} {\bibinfo {author} {\bibfnamefont {J.}~\bibnamefont {del
  Valle}}, \bibinfo {author} {\bibfnamefont {P.}~\bibnamefont {Salev}},
  \bibinfo {author} {\bibfnamefont {F.}~\bibnamefont {Tesler}}, \bibinfo
  {author} {\bibfnamefont {N.~M.}\ \bibnamefont {Vargas}}, \bibinfo {author}
  {\bibfnamefont {Y.}~\bibnamefont {Kalcheim}}, \bibinfo {author}
  {\bibfnamefont {P.}~\bibnamefont {Wang}}, \bibinfo {author} {\bibfnamefont
  {J.}~\bibnamefont {Trastoy}}, \bibinfo {author} {\bibfnamefont {M.-H.}\
  \bibnamefont {Lee}}, \bibinfo {author} {\bibfnamefont {G.}~\bibnamefont
  {Kassabian}}, \bibinfo {author} {\bibfnamefont {J.~G.}\ \bibnamefont
  {Ram\'{i}rez}}, \bibinfo {author} {\bibfnamefont {M.~J.}\ \bibnamefont
  {Rozenberg}},\ and\ \bibinfo {author} {\bibfnamefont {I.~K.}\ \bibnamefont
  {Schuller}},\ }\bibfield  {title} {\bibinfo {title} {Subthreshold firing in
  {M}ott nanodevices},\ }\href {https://doi.org/10.1038/s41586-019-1159-6}
  {\bibfield  {journal} {\bibinfo  {journal} {Nature}\ }\textbf {\bibinfo
  {volume} {569}},\ \bibinfo {pages} {388} (\bibinfo {year}
  {2019})}\BibitemShut {NoStop}%
\bibitem [{\citenamefont {Aonuma}\ \emph {et~al.}(1993)\citenamefont {Aonuma},
  \citenamefont {Sawa}, \citenamefont {Okano}, \citenamefont {Kato},\ and\
  \citenamefont {Kobayashi}}]{Aonuma1993}%
  \BibitemOpen
  \bibfield  {author} {\bibinfo {author} {\bibfnamefont {S.}~\bibnamefont
  {Aonuma}}, \bibinfo {author} {\bibfnamefont {H.}~\bibnamefont {Sawa}},
  \bibinfo {author} {\bibfnamefont {Y.}~\bibnamefont {Okano}}, \bibinfo
  {author} {\bibfnamefont {R.}~\bibnamefont {Kato}},\ and\ \bibinfo {author}
  {\bibfnamefont {H.}~\bibnamefont {Kobayashi}},\ }\bibfield  {title} {\bibinfo
  {title} {Synthesis of {DM}e-{DCNQI}-d$_{7}$ and deuterium-induced
  metal-insulator transition of ({DM}e-{DCNQI}-d$_{7}$)$_{2}${C}u},\ }\href
  {https://doi.org/10.1016/0379-6779(93)91115-I} {\bibfield  {journal}
  {\bibinfo  {journal} {Synthetic Metals}\ }\textbf {\bibinfo {volume} {58}},\
  \bibinfo {pages} {29} (\bibinfo {year} {1993})}\BibitemShut {NoStop}%
\bibitem [{\citenamefont {Aonuma}\ \emph {et~al.}(1995)\citenamefont {Aonuma},
  \citenamefont {Sawa},\ and\ \citenamefont {Kato}}]{Aonuma1995}%
  \BibitemOpen
  \bibfield  {author} {\bibinfo {author} {\bibfnamefont {S.}~\bibnamefont
  {Aonuma}}, \bibinfo {author} {\bibfnamefont {H.}~\bibnamefont {Sawa}},\ and\
  \bibinfo {author} {\bibfnamefont {R.}~\bibnamefont {Kato}},\ }\bibfield
  {title} {\bibinfo {title} {Chemical pressure effect by selective deuteriation
  in the molecular-based conductor,
  2,5-dimethyl-{N},{N}'-dicyano-p-benzoquinone immine-copper salt,
  ({DM}e-{DCNQI})$_{2}${C}u},\ }\href {https://doi.org/10.1039/P29950001541}
  {\bibfield  {journal} {\bibinfo  {journal} {Journal of the Chemical Society,
  Perkin Transactions 2}\ ,\ \bibinfo {pages} {1541}} (\bibinfo {year}
  {1995})}\BibitemShut {NoStop}%
\bibitem [{\citenamefont {Kato}(2000)}]{Kato2000}%
  \BibitemOpen
  \bibfield  {author} {\bibinfo {author} {\bibfnamefont {R.}~\bibnamefont
  {Kato}},\ }\bibfield  {title} {\bibinfo {title} {Conductive {C}opper {S}alts
  of 2,5-{D}isubstituted {N}, {N}'-{D}icyanobenzoquinonediimines ({DCNQI}s):
  {S}tructural and {P}hysical {P}roperties},\ }\href
  {https://doi.org/10.1246/bcsj.73.515} {\bibfield  {journal} {\bibinfo
  {journal} {Bulletin of the Chemical Society of Japan}\ }\textbf {\bibinfo
  {volume} {73}},\ \bibinfo {pages} {515} (\bibinfo {year} {2000})}\BibitemShut
  {NoStop}%
\bibitem [{\citenamefont {Fukuyama}(1992)}]{Fukuyama1992}%
  \BibitemOpen
  \bibfield  {author} {\bibinfo {author} {\bibfnamefont {H.}~\bibnamefont
  {Fukuyama}},\ }\bibfield  {title} {\bibinfo {title} {({DCNQI})$_{2}${C}u: A
  {L}uttinger-{P}eierls {S}ystem},\ }\href
  {https://doi.org/10.1143/JPSJ.61.3452} {\bibfield  {journal} {\bibinfo
  {journal} {Journal of the Physical Society of Japan}\ }\textbf {\bibinfo
  {volume} {61}},\ \bibinfo {pages} {3452} (\bibinfo {year}
  {1992})}\BibitemShut {NoStop}%
\bibitem [{\citenamefont {Fukuyama}(2006)}]{Fukuyama2006}%
  \BibitemOpen
  \bibfield  {author} {\bibinfo {author} {\bibfnamefont {H.}~\bibnamefont
  {Fukuyama}},\ }\bibfield  {title} {\bibinfo {title} {Physics of {M}olecular
  {C}onductors},\ }\href {https://doi.org/10.1143/JPSJ.75.051001} {\bibfield
  {journal} {\bibinfo  {journal} {Journal of the Physical Society of Japan}\
  }\textbf {\bibinfo {volume} {75}},\ \bibinfo {pages} {051001} (\bibinfo
  {year} {2006})}\BibitemShut {NoStop}%
\bibitem [{\citenamefont {Ridley}(1963)}]{Ridley1963}%
  \BibitemOpen
  \bibfield  {author} {\bibinfo {author} {\bibfnamefont {B.~K.}\ \bibnamefont
  {Ridley}},\ }\bibfield  {title} {\bibinfo {title} {Specific {N}egative
  {R}esistance in {S}olids},\ }\href
  {https://doi.org/10.1088/0370-1328/82/6/315} {\bibfield  {journal} {\bibinfo
  {journal} {Proceedings of the Physical Society}\ }\textbf {\bibinfo {volume}
  {82}},\ \bibinfo {pages} {954} (\bibinfo {year} {1963})}\BibitemShut
  {NoStop}%
\bibitem [{\citenamefont {Kumar}\ and\ \citenamefont
  {Williams}(2018)}]{Kumar2018}%
  \BibitemOpen
  \bibfield  {author} {\bibinfo {author} {\bibfnamefont {S.}~\bibnamefont
  {Kumar}}\ and\ \bibinfo {author} {\bibfnamefont {R.~S.}\ \bibnamefont
  {Williams}},\ }\bibfield  {title} {\bibinfo {title} {Separation of current
  density and electric field domains caused by nonlinear electronic
  instabilities},\ }\href {https://doi.org/10.1038/s41467-018-04452-w}
  {\bibfield  {journal} {\bibinfo  {journal} {Nature Communications}\ }\textbf
  {\bibinfo {volume} {9}},\ \bibinfo {pages} {2030} (\bibinfo {year}
  {2018})}\BibitemShut {NoStop}%
\bibitem [{\citenamefont {Goodwill}\ \emph {et~al.}(2019)\citenamefont
  {Goodwill}, \citenamefont {Ramer}, \citenamefont {Li}, \citenamefont
  {Hoskins}, \citenamefont {Pavlidis}, \citenamefont {McClelland},
  \citenamefont {Centrone}, \citenamefont {Bain},\ and\ \citenamefont
  {Skowronski}}]{Goodwill2019}%
  \BibitemOpen
  \bibfield  {author} {\bibinfo {author} {\bibfnamefont {J.~M.}\ \bibnamefont
  {Goodwill}}, \bibinfo {author} {\bibfnamefont {G.}~\bibnamefont {Ramer}},
  \bibinfo {author} {\bibfnamefont {D.}~\bibnamefont {Li}}, \bibinfo {author}
  {\bibfnamefont {B.~D.}\ \bibnamefont {Hoskins}}, \bibinfo {author}
  {\bibfnamefont {G.}~\bibnamefont {Pavlidis}}, \bibinfo {author}
  {\bibfnamefont {J.~J.}\ \bibnamefont {McClelland}}, \bibinfo {author}
  {\bibfnamefont {A.}~\bibnamefont {Centrone}}, \bibinfo {author}
  {\bibfnamefont {J.~A.}\ \bibnamefont {Bain}},\ and\ \bibinfo {author}
  {\bibfnamefont {M.}~\bibnamefont {Skowronski}},\ }\bibfield  {title}
  {\bibinfo {title} {Spontaneous current constriction in threshold switching
  devices},\ }\href {https://doi.org/10.1038/s41467-019-09679-9} {\bibfield
  {journal} {\bibinfo  {journal} {Nature Communications}\ }\textbf {\bibinfo
  {volume} {10}},\ \bibinfo {pages} {1628} (\bibinfo {year}
  {2019})}\BibitemShut {NoStop}%
\bibitem [{\citenamefont {Kumar}\ \emph
  {et~al.}(2017{\natexlab{b}})\citenamefont {Kumar}, \citenamefont {Wang},
  \citenamefont {Davila}, \citenamefont {Kumari}, \citenamefont {Norris},
  \citenamefont {Huang}, \citenamefont {Strachan}, \citenamefont {Vine},
  \citenamefont {Kilcoyne}, \citenamefont {Nishi},\ and\ \citenamefont
  {Williams}}]{Kumar2017_2}%
  \BibitemOpen
  \bibfield  {author} {\bibinfo {author} {\bibfnamefont {S.}~\bibnamefont
  {Kumar}}, \bibinfo {author} {\bibfnamefont {Z.}~\bibnamefont {Wang}},
  \bibinfo {author} {\bibfnamefont {N.}~\bibnamefont {Davila}}, \bibinfo
  {author} {\bibfnamefont {N.}~\bibnamefont {Kumari}}, \bibinfo {author}
  {\bibfnamefont {K.~J.}\ \bibnamefont {Norris}}, \bibinfo {author}
  {\bibfnamefont {X.}~\bibnamefont {Huang}}, \bibinfo {author} {\bibfnamefont
  {J.~P.}\ \bibnamefont {Strachan}}, \bibinfo {author} {\bibfnamefont
  {D.}~\bibnamefont {Vine}}, \bibinfo {author} {\bibfnamefont {A.~D.}\
  \bibnamefont {Kilcoyne}}, \bibinfo {author} {\bibfnamefont {Y.}~\bibnamefont
  {Nishi}},\ and\ \bibinfo {author} {\bibfnamefont {R.~S.}\ \bibnamefont
  {Williams}},\ }\bibfield  {title} {\bibinfo {title} {Physical origins of
  current and temperature controlled negative differential resistances in
  {N}b{O}$_{2}$},\ }\href {https://doi.org/10.1038/s41467-017-00773-4}
  {\bibfield  {journal} {\bibinfo  {journal} {Nature Communications}\ }\textbf
  {\bibinfo {volume} {8}},\ \bibinfo {pages} {658} (\bibinfo {year}
  {2017}{\natexlab{b}})}\BibitemShut {NoStop}%
\bibitem [{\citenamefont {Mott}(1990)}]{NFMott1990}%
  \BibitemOpen
  \bibfield  {author} {\bibinfo {author} {\bibfnamefont {N.~F.}\ \bibnamefont
  {Mott}},\ }\href {https://doi.org/10.1201/b12795} {\emph {\bibinfo {title}
  {Metal-Insulator Transitions}}}\ (\bibinfo  {publisher} {Taylor \& Francis,
  London},\ \bibinfo {year} {1990})\BibitemShut {NoStop}%
\bibitem [{\citenamefont {Hiraki}\ \emph {et~al.}(1995)\citenamefont {Hiraki},
  \citenamefont {Kobayashi}, \citenamefont {Nakamura}, \citenamefont
  {Takahashi}, \citenamefont {Aonuma}, \citenamefont {Sawa}, \citenamefont
  {Kato},\ and\ \citenamefont {Kobayashi}}]{Hiraki1995}%
  \BibitemOpen
  \bibfield  {author} {\bibinfo {author} {\bibfnamefont {K.}~\bibnamefont
  {Hiraki}}, \bibinfo {author} {\bibfnamefont {Y.}~\bibnamefont {Kobayashi}},
  \bibinfo {author} {\bibfnamefont {T.}~\bibnamefont {Nakamura}}, \bibinfo
  {author} {\bibfnamefont {T.}~\bibnamefont {Takahashi}}, \bibinfo {author}
  {\bibfnamefont {S.}~\bibnamefont {Aonuma}}, \bibinfo {author} {\bibfnamefont
  {H.}~\bibnamefont {Sawa}}, \bibinfo {author} {\bibfnamefont {R.}~\bibnamefont
  {Kato}},\ and\ \bibinfo {author} {\bibfnamefont {H.}~\bibnamefont
  {Kobayashi}},\ }\bibfield  {title} {\bibinfo {title} {Magnetic {S}tructure in
  the {A}ntiferromagnetic {S}tate of the {O}rganic {C}onductor,
  ({DM}e-{DCNQI}[3,3:1]d$_{7}$)$_{2}${C}u:$^{1}${H}-{NMR} {A}nalysis},\ }\href
  {https://doi.org/10.1143/JPSJ.64.2203} {\bibfield  {journal} {\bibinfo
  {journal} {Journal of the Physical Society of Japan}\ }\textbf {\bibinfo
  {volume} {64}},\ \bibinfo {pages} {2203} (\bibinfo {year}
  {1995})}\BibitemShut {NoStop}%
\bibitem [{Sup()}]{Supplementary}%
  \BibitemOpen
  \href@noop {} {\ }\bibinfo {note} {See Supplemental Material appended to this 
  manuscript for the supporting results and discussion}\BibitemShut
  {NoStop}%
\end{thebibliography}
\end{document}